\documentclass[fleqn,10pt]{wlscirep}
\usepackage{geometry}

\usepackage{amsmath}

\makeatletter
\newcommand{\inlineeq}[2]{%
  \refstepcounter{equation}
  \label{#1}
  \ensuremath{#2}\ \textnormal{(\theequation)}%
}
\makeatother

\usepackage[normalem]{ulem}
\usepackage[switch]{lineno}

\usepackage{graphicx}%
\usepackage{multirow}%
\usepackage{amsmath,amssymb,amsfonts}%
\usepackage{amsthm}%
\usepackage{mathrsfs}%
\usepackage[title]{appendix}%
\usepackage{xcolor}%
\usepackage{textcomp}%
\usepackage{manyfoot}%
\usepackage{booktabs}%
\usepackage{algorithm}%
\usepackage{algorithmicx}%
\usepackage{algpseudocode}%
\usepackage{listings}%
\usepackage{enumitem}
\usepackage{amsmath}
\usepackage{subcaption}

\usepackage{cleveref}
\begin{document}
\title{\vspace{2cm} The spatial organization of wind turbine wakes}

\author[1]{Janka Lengyel$^{*}$}
\author[2]{{St\'ephane G.} {Roux}}
\author[2]{{Patrice} {Abry}}
\author[3]{{Norman} {Wildmann}}
\author[3]{{Julia} {Menken}}
\author[4]{{Olivier} {Bonin}}
\author[1]{{Jan} {Friedrich}}

\affil{%
Carl von Ossietzky Universität Oldenburg, School of Mathematics and Science, Institute of Physics, ForWind - Center for Wind Energy Research, Oldenburg 26129, Germany;
  $^2$ CNRS, ENS de Lyon, Laboratoire de Physique, Lyon 69007, France; %
  $^3$ Deutsches Zentrum für Luft- und Raumfahrt, Institut für Physik der Atmosphäre, Oberpfaffenhofen, Germany; %
  $^4$ Univ. Gustave Eiffel, École des Ponts, LVMT, Champs-sur-Marne 77420, France, %
  $^{*}$ {janka.lengyel@uni-oldenburg.de}%
}

\begin{abstract}
    
Wind turbine wakes play a central role in determining wind farm performance, yet their spatial organization remains only partially understood. Here, we apply a spatially localized multifractal analysis to quantify the strength of dependencies (local roughness) and extreme velocity fluctuations (local intermittency) in turbine wakes, and relate these properties to established metrics in wind energy research. Using two-dimensional nacelle-mounted LiDAR plan-position-indicator scans, we extract scale-invariant features that enable systematic comparisons across the wake without requiring time-resolved data. Designed to robustly handle irregular sampling, our analysis yields four main findings:
\emph{i.)} Four distinct wake zones are identified, each exhibiting unique patterns of roughness and intermittency.
\emph{ii.)} Coherent, strongly correlated patches emerge 2 to 5 rotor diameters $D$ downstream, with intermittency strengthening periodically at multiple $D$ positions and along the wake-free-flow interface.
\emph{iii.)} The classical ``intermittency ring'' is consequently redefined as a set of localized ``intermittency bubbles'',
\emph{iv.)} which interact dynamically with the ambient atmosphere through an inverse energy cascade, transferring energy from small to large scales. These findings, supported by concurrent cup anemometer observations under free-inflow conditions, demonstrate that local multifractal analysis provides a robust and cost-effective diagnostic framework for wake characterization and wake-model validation, with direct relevance for wind-farm design and control.
\end{abstract}
\maketitle
\vspace{-0.4cm}
\section*{Main}

Renewable energy, and wind power in particular, is increasingly integrated into the framework of Nationally Determined Contributions and national net-zero roadmaps \cite{IEA_NetZero2023,UNFCCC_2025_NDC_Synthesis}. Accordingly, wind energy has emerged as a key technology in global decarbonization pathways, with many countries adopting ambitious targets to boost its share in the energy mix \cite{IEA_WEO2024}. Meanwhile, the rapid increase in wind turbine size in new deployments and repowering projects, coupled with the scarcity of sites with favourable wind resources, has resulted in an increasing fraction of turbines operating under wake-influenced inflow conditions \cite{veers2023grand}. Against this backdrop, the present study sets out to investigate the \textit{intrinsic spatial organization of turbine wakes}. While wake research has traditionally focused on aggregated variables such as mean velocity deficit and turbulence intensity, the multiscale spatial dynamics of instantaneous wake structures has received comparatively little attention. This is despite the fact that a fundamental understanding of the spatio-temporal organization of the wake developing behind wind turbines is essential, as wake effects directly influence both power output \cite{Stevens2017} and fatigue loads on the turbine \cite{Shaler2023}. At the wind-farm scale, wake interactions can significantly reduce overall efficiency and further complexify layout optimization, making wake research central to improving the performance of wind energy systems~\cite{porte2020wind}. Our aim is therefore twofold: \textit{i.)} to characterize the heterogeneous morphology and dynamics of real-world turbine wakes, including their localized interactions with the surrounding atmospheric boundary layer - features that are often difficult to reproduce in idealized laboratory or numerical settings and yet are critical for understanding fundamental wake dynamics such as wake recovery and meandering; and \textit{ii.)} to develop a more differentiated delineation of wake zones (beyond the conventional near- and far-wake distinction), providing increased spatial detail and structural discrimination that may not be attainable with conventional wind-energy metrics.

\section*{Spatially localized feature extraction in non-homogeneous LiDAR data}
Building on recent wind-tunnel and large-eddy simulation studies resolving fine-scale wake structure \cite{neunaber2020distinct,schottler2018wind,Wu2024,Zhang2023, scott2023evolution}, we analyze turbine wakes using a recently developed spatially localized multifractal (LMF) framework \cite{Lengyel2022_localMultifractalityUrban,Lengyel2024_LomPy,lengyel2023roughness,lengyel2025bivariate}. The approach extends conventional single-scale wind-energy metrics by jointly quantifying local spatial correlations (roughness) and extreme fluctuations (intermittency) across scales. We apply LMF to two-dimensional nacelle-mounted LiDAR plan-position-indicator (PPI) scans (hereafter scans), \cite{DLR_WiValdi} collected downstream of a utility-scale turbine with a rotor diameter of $D=115.7$ m under real atmospheric conditions (see Methods and Extended Data (ED) Fig.~\ref{fig:ext_cup_alpha} for full details on the dataset and its challenges). The analysis operates on the measured radial (or line-of-sight) velocity~\cite{stoevesandt2022handbook} ($\mathrm{m\,s^{-1}}$), denoted as $\kappa_e$, which is irregularly sampled at locations $e(x_e,y_e)$. It produces spatially resolved fields that can be directly linked to classical wind-energy science (CWES) metrics; All CWES and LMF parameters are estimated on an equidistant grid $g(x_g,y_g)$ with a grid spacing of $15$ m, spanning the full data extent, as illustrated in ED Fig.~\ref{fig:ext_method}. The original-data/grid connectivity is defined by the weight $w_{g,e}=1$ if $d_{g,e}<L$, and $0$ otherwise, with $d_{g,e}=\sqrt{(x_e-x_g)^2+(y_e-y_g)^2}$, followed by normalization such that $\sum_e w_{g,e}=1$. In the CWES context, the localized mean, standard deviation, velocity deficit, and turbulence intensity at grid point $g$ and scale $L$ are defined compactly as
\begin{center}
\small
\inlineeq{eqn:I}{\mu(g,L)=\sum_e w_{g,e}\kappa_e,}
\qquad
\inlineeq{eqn:sigma}{\sigma(g,L)=\sqrt{\sum_e w_{g,e}\bigl(\kappa_e-\mu(g,L)\bigr)^2},}
\qquad
\inlineeq{eqn:VD}{\mathrm{VD}(g,L)=\frac{\mu(g,L)}{\max_g(\mu(g,L))},}
\qquad
\inlineeq{eqn:TI}{\mathrm{TI}(g,L)=\frac{\sigma(g,L)}{\mu(g,L)}.}
\end{center}
The core quantities of local multifractal analysis are the wavelet-like multiresolution quantities $D(e,r)$, centered on $e$ and evaluated across length scales $r$, and their corresponding $p$-leaders $T(e,r,p)$ \cite{wendt2007bootstrap,wendt2009wavelet} defined according to 
\begin{center}
\small
\inlineeq{eqn:sb_mark}{\mu(e,r) = \frac{1}{N(e,r)} \sum_{e' \,:\, d_{e,e'} \leq r} \kappa_{e'},}
\qquad
\inlineeq{eqn:wt_hat_mark}{D(e,r) = \mu(e,r) - \mu(e,\sqrt{2}r),}
\qquad
\inlineeq{eqn:leader}{T(e,r,p) = 
    \left(
        \frac{1}{N(e,r)}  \sum_{\substack{e' \,:\, d_{e,e'} \leq r }}
        |D(e',r)|^p 
    \right)^{1/p},}
\end{center}
where $d_{e,e'} = \sqrt{(x_e - x_{e'})^2 + (y_e - y_{e'})^2}$, and $N(e,r)$ denotes the number of points $e$ within $r$ as defined in Methods. From these, the first- and second-order log-cumulants are estimated, and the associated scaling exponents are derived as
\begin{center}
\small
\inlineeq{eq:C1}{C_1(g,r,L) = \sum_{e} w_{g,e} \, \log |D(e,r)| 
    \sim c_1(g,L) \, \log(r) + k_1(g,L),}
\inlineeq{eq:C2}{C_2(g,r,L) = \sum_{e} w_{g,e} 
    \left[\log |T(e,r,p)| - C_1(g,r,L)\right]^2 
    \sim_{r\to0} c_2(g,L) \, \log(r) + k_2(g,L).}
\end{center}
The exponent $c_1(g,L)$ captures the strength of local spatial dependence or \emph{local roughness}, while $c_2(g,L)$ quantifies localized non-Gaussian fluctuations or \emph{local intermittency}; together, their spatial organization serves here to yield a structurally resolved description of the wake morphology. Full LMF methodological details are also provided in the Methods. The multifractal analysis is performed over scales $r=55$--$170$ m ($0.47$--$1.47 D$), corresponding to the most robust scaling regime observed in the data (ED Fig.~\ref{fig:ext_method}b,c). The smallest scale lies just above the LiDAR pulse length ($50~\mathrm{m}$), while the largest scale defines the local environment size, $L = 170~\mathrm{m}$, which is used consistently across all classical wind-energy metrics. Our findings based on the LiDAR data are complemented and supported by concurrent cup-anemometer observations under free-inflow conditions, as presented in the final section of this paper. 
\section*{Instantaneous wake dynamics}
This and the following section focuses exclusively on the results obtained from the nacelle LiDAR data. For clarity, we distinguish between three aggregation levels of the results: \textit{(1) Instantaneous}: results derived from a single scan downstream of the turbine. Each here-used LiDAR scan requires $\approx$ 5.83 s, and the analyzed window is chosen to minimize distortions due to intra-scan temporal lag. \textit{(2) Short-term averaging}: results averaged over ten scans (approximately five minutes of elapsed time), providing robustness while still retaining as much local variability as possible. \textit{(3) Long-term averaging}: results averaged over $222$ scans (or approximately two hours of elapsed time) to obtain stronger statistical reliability.  Correlation values between all extracted CWES and LMF fields at each aggregation level are reported in ED Table 1. It must also be noted here  that the mean coefficients of determination $R^2$ across the $222$ analyzed scans and all estimation sites are $R^2(c_1(g,L))=0.8397$ and $R^2(c_2(g,L))=0.7986$, indicating a rather robust analysis performance in the context of irregularly sampled point processes. \par
\begin{figure}[!t]
    \makebox[\textwidth][c]{%
        \begin{minipage}{1.\linewidth}
            \centering
            \includegraphics[width=\linewidth]{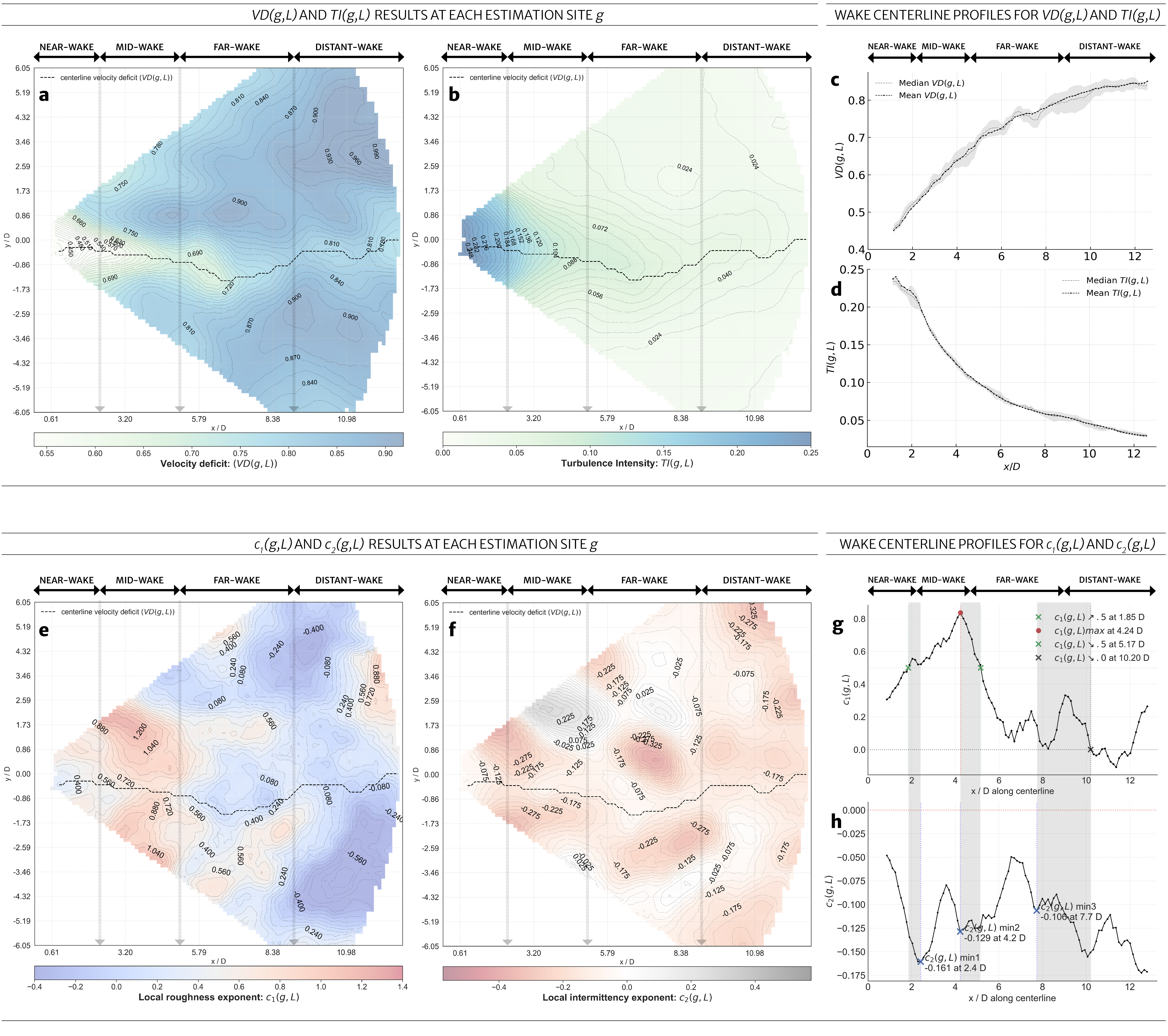}
        \end{minipage}
    }
    \caption{\captiontitle{Top: Instantaneous results of the CWES metrics}. Three important remarks apply here: \emph{i.)} The velocity deficit $VD(g,L)$ shown in (a) was used to extract the wake centerline path using the computation method detailed in the Methods. \emph{ii.)} Velocity deficit (a) and turbulence intensity $TI(g,L)$ (b) convey distinct physical information in theory, yet in practice the two fields are highly anti-correlated ($\rho = -0.78$) at resolution $L$. \emph{iii.)} The wake zones derived from the LMF parameters (g,h), when projected onto the CWES maps (a,b), exhibit clear visual consistency with the underlying flow structure. Panels (c,d) show the \emph{short-term} centerline statistics for the CWES metrics. \captiontitle{Bottom: Instantaneous results of the LMF parameters}. The estimates at each site $g$ for $c_1(g,L)$ (e) and $c_2(g,L)$ (f) exhibit distinct patterns from each other ($\rho = -0.01$) and from those of $VD(g,L)$ and $TI(g,L)$ in (a,b) - for all pairwise cross-correlations see ED Table 1. The corresponding centerline profiles (g,h) also differ markedly in their tendencies. Their landmarks are used jointly to interpret the wake structure and to delineate the distinct wake zones. The inflow wind speed at 00:07 is $6.19~\mathrm{m\,s^{-1}}$ (ED Fig.~\ref{fig:mfcup_48}.a).}
    \label{fig:main_c1c2}
\end{figure}

 During the measurement period, the vertical wind-speed profile exhibited shear exponents of $0.27$--$0.33$ (ED Fig.~\ref{fig:ext_cup_alpha}), consistent with stable nocturnal boundary-layer conditions. This stable regime was selected to \textit{establish a robust reference} for extracting localized higher-order structural parameters from non-uniformly sampled LiDAR data, with the goal of providing a foundation for future work rather than a longitudinal analysis across diverse atmospheric conditions. A second instantaneous and long-term case study with higher wind speeds is included in the Extended Data (ED Fig.~\ref{fig:ext_second_cs}) to provide additional support. 
\paragraph{Classical wind energy science} Fig.~\ref{fig:main_c1c2}a,b,c,d summarizes the classical characterization of the wake based on velocity deficit $VD(g,L)$ and turbulence intensity $TI(g,L)$. The velocity-deficit field (a) shows the expected reduction in wind speed downstream of the turbine, with the  most substantial deficits located near the rotor and a gradual recovery farther downstream. The pattern is smooth and coherent, in agreement with established Gaussian wake descriptions and commonly used engineering models \cite{Bastankhah2016,Burton2011}. However, owing to the local smoothing over the environmental scale L, the velocity deficit exhibits a single Gaussian peak rather than the double-Gaussian structure more typically observed, for example, in wind-tunnel experiments \cite{neunaber2020distinct}. The $VD(g,L)$ field is also used to extract the wake centerline (see Methods). This centerline is then superimposed on all other CWES and LMF parameter fields to compute centerline-based statistics. The turbulence-intensity field (b) exhibits a similarly smooth structure: a concentrated region of elevated $TI(g,L)$ immediately behind the turbine, which steadily decreases with downstream distance. This behaviour aligns with field observations and classical parameterizations such as the Frandsen wake-added turbulence model \cite{Frandsen2007}. Together, these fields provide a coherent yet relatively simple picture of wake behaviour, dominated by a single description of uninterrupted change as a function of downstream distance (supported also by the short-term centerline results in Fig.~\ref{fig:main_c1c2}c and Fig.~\ref{fig:main_c1c2}d). 
\paragraph{Local multifractal analysis} In sharp contrast, the LMF exponents shown in Fig.~\ref{fig:main_c1c2}e,f,g,h reveal that the spatial organization of the wake may be far more heterogeneous than suggested by the CWES parameters above. The fractal and multifractal fields in Fig.~\ref{fig:main_c1c2}e,f exhibit pronounced spatial variability at first glance, with no clear correspondence to regions of most substantial velocity deficit \(VD(g, L)\) or highest turbulence intensity \(TI(g, L)\) (see ED Table~1a for the exact correlation values).  
\subparagraph{Local roughness} Fig.~\ref{fig:main_c1c2}e shows the extracted roughness exponent $c_1(g, L)$ at every estimation site $g$. Larger values - represented by darker red colors - indicate stronger spatial correlations in the 2D velocity field. The divergent colorbar switches at $c_1(g, L) = 0.5$, marking the transition between short- ($c_1(g, L)< 0.5$) and long-range correlated processes, a distinction that will play an important role in our interpretation. In Fig.~\ref{fig:main_c1c2}e, a clear zone of high coherence $0.5<c_1(g, L)< 1.0$ emerges between $1.9D$ and $5.2D$. This is confirmed by the streamwise profile of $c_1(g, L)$ (Fig.~\ref{fig:main_c1c2}c) extracted along the wake centerline: after an initial region with only moderately correlated behavior ($0.25 < c_1(g, L) < 0.5$), the profile enters the aforementioned high-coherence zone, before transitioning again to a rougher signal further downstream ($5.2 < x/D < 10.2$). Eventually, $c_1(g, L)$ becomes negative, indicating fluctuations that resemble a highly irregular, yet still non-Gaussian noise-like process, which will be discussed in the next paragraph and in the cup-anemometer results section. 

\subparagraph{Local intermittency} The second extracted feature, $c_2(g, L)$, shown in Fig.~\ref{fig:main_c1c2}f, is indicative of the multifractality of the process, or equivalently, the deviations from Gaussian behavior at each grid point $g$. The darker the red color - or the more negative the value - the stronger the inferred intermittency and the higher the probability of extreme fluctuations emerging toward smaller scales. Three dominant trends can be identified. \textit{i.)} First, intermittency intensifies along the wake/free-flow interface, a phenomenon also observed in previous wind-tunnel experiments \cite{neunaber2020distinct, schottler2018wind}, where it was termed the \emph{intermittency ring}. Notably, those results were based on one-dimensional, temporally resolved hot-wire measurements. \textit{ii.)} Second, elevated intermittency emerges periodically at several downstream positions (multiples of $D$) within the wake core, as further confirmed by the centerline profile of $c_2(g, L)$ in Fig.~\ref{fig:main_c1c2}h: the first minimum occurs near $2.4D$, after which intermittency partially recovers (i.e., $c_2(g,L)$ becomes less negative) before collapsing again at a second minimum around $4.2D$. This pattern repeats one or two more times farther downstream, depending on the LiDAR scan and therefore the given instantaneous atmospheric condition. Because these intermittency cycles are central to the development of key wake-dynamics features in this study, three additional instantaneous $c_2(g,L)$ fields are presented in ED Fig.~\ref{fig:ext_int_cycles} to further illustrate their recurrent behavior. Such modulation may be consistent with \textit{wake meandering}, whose low-frequency lateral motion produces alternating zones of enhanced and reduced turbulence activity \cite{heisel2018spectral, Foti2018, li2022onset}. \textit{iii.)} Third, between the layer of higher intermittency along the wake/free-flow interface and the ambient-atmosphere intermittency, a region of grey colors appears, indicating vanishing or even positive values of $c_2(g, L)$. 
It must be stressed that positive values of $c_2(g, L)$ are uncommon in classical multifractal analysis and can arise from estimation bias. Notably however, the $c_2(g, L)>0$ zone persists in the extended long-term evaluation shown in Fig.~\ref{fig:main_c1c2_long}b, suggesting that in this setting it may reflect a genuine property of the wake dynamics. This behaviour implies that the probability of extreme fluctuations grows with scale, consistent with inverse-cascade processes that transport energy from smaller toward larger motions. Comparable upscale transfer has been documented in shear-driven and quasi-two-dimensional turbulent systems~\cite{AlexakisBiferale2018, pumir2014redistribution}. A possible physical mechanism underlying the emergence of an inverse-cascade region is \emph{vortex thinning}~\cite{chen2006physical,friedrich2013generalized}. In this process, axisymmetric vortices - such as paired tip vortices in the near wake~\cite{biswas2024effect} - deform under their mutual strain field and are stretched into elliptic shapes. This thinning induces relative attractive motion between the vortices, reducing the kinetic energy at small scales and thereby promoting energy transfer toward larger scales.

\section*{Persistent wake features and the four extracted wake zones}

\begin{figure}[t!]
    \makebox[\textwidth][c]{%
        \begin{minipage}{1\linewidth}
            \centering
            \includegraphics[width=\linewidth]{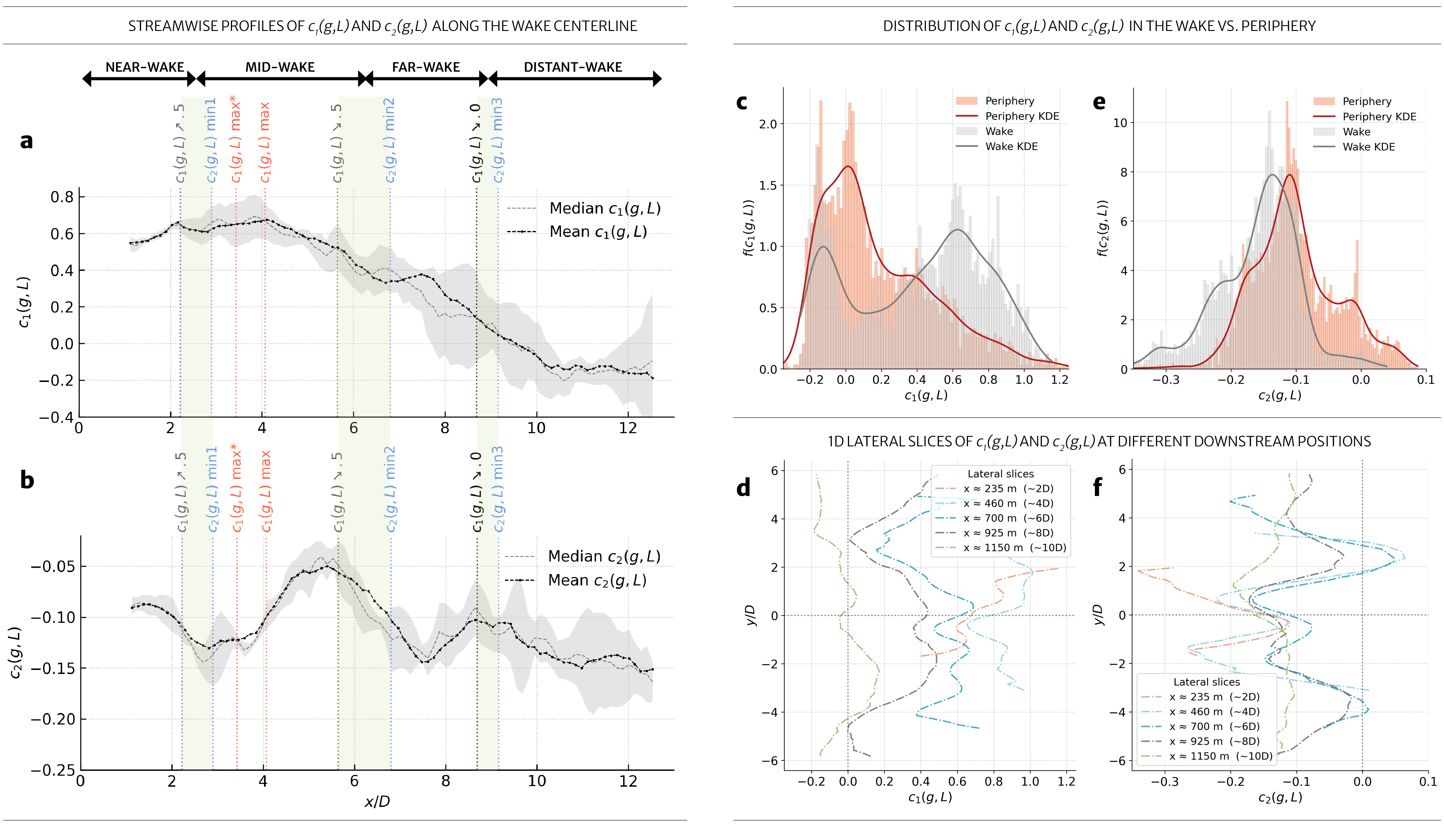}
        \end{minipage}
    }
    \caption{\captiontitle{Short-term results}. The LMF wake centerline statistics (a,b) display key trends and landmarks that are used here to delineate the wake zones: $c_1(g, L) \nearrow .5$ (the first $0.5$ crossing), $c_1(g, L) \searrow .5$ the first $0.5$ crossing after $c_1(g, L)$ reaches its maximum $c_1(g, L)\,\max$, and the recurring negative peaks $c_2(g, L)\,\min$.  Note that the landmarks shown here are computed as averages of the individual scan-wise $x/D$ landmark locations (see Fig.\ref{fig:main_c1c2}g and Fig.\ref{fig:main_c1c2}h), rather than being derived from the averaged centerline. For illustration, we also include $c_1(g,L){\max}^*$, defined as the median of these extracted locations, in contrast to the mean value used in all other analyses: $c_1(g,L){\max}$, which is also shown. Differences between the wake and periphery histograms of the LMF indicators (c,e), together with the lateral slices at five downstream positions (d,f), help to identify the dominant structures within the wake zones. The corresponding LMF fields are displayed in ED Fig. \ref{fig:ext_c1c2_shortterm}. The average inflow wind speed for the short-term window at hub height is $6.21~\mathrm{m\,s^{-1}}$ (ED Fig.~\ref{fig:mfcup_48}a).}
    \label{fig:short_term_results}
\end{figure}

\paragraph{Short-term results} The short-term aggregation level was selected because it provides sufficient robustness for interpretation while still preserving local variability, including the inherent meandering motion and other small-scale dynamical features of the wake. Synthesizing these observations, we delineate four characteristic wake regions; these are based on a \textit{joint analysis} of the key crossings, local peaks, and minima in the LMF centreline statistics, as described in the caption of Fig.~\ref{fig:short_term_results}. Additionally, for further support, we adopt a wake–periphery distinction, defining the wake core as approximately $-2 < y/D < 2$ based on Fig.~\ref{fig:main_c1c2}a and ED Fig.~\ref{fig:ext_TI_VD_long}a, with the periphery comprising the remaining estimation sites. 

\subparagraph{1. Near-wake, transition to next zone around $2<x/D<3$}
In Fig.~\ref{fig:short_term_results}a, $c_1(g, L)$ increases steadily while still indicating relatively rough behaviour, whereas $c_2(g, L)$ becomes progressively more negative. This trend persists until $c_1(g, L)$ crosses its first 0.5 threshold - marking the transition from short- to long-range correlations -  and $c_2(g, L)$ reaches its first minimum, signaling strong localized deviations from Gaussianity; together, these two ``landmarks'' jointly delineate the end of the near-wake zone. The lateral (cross-stream) profiles at hub height in Fig.~\ref{fig:short_term_results}f at $\sim 2D$ show the strongest centerline dip in $c_2(g,L)$, together with a clear lateral asymmetry in both $c_1(g,L)$ and $c_2(g,L)$ (Fig.~\ref{fig:short_term_results}d and Fig.~\ref{fig:short_term_results}f). Such early-stage asymmetries in lateral wake structure have previously been identified as precursors of large-scale meandering motions \cite{Medici2006}. Finally, it is within the near-wake region, that the CWES metrics display their most pronounced modulation, consistent with the behaviour discussed above.
\subparagraph{2. Mid-wake, transition to next zone around $5<x/D<6.5$} According to the centerlines in Fig.~\ref{fig:short_term_results}a and Fig.~\ref{fig:short_term_results}b, this zone is characterized by pronounced spatial correlations and a complete intermittency cycle. Within this region of $0.5<c_1(g,L)<1.0$ values, the wake histogram (Fig.~\ref{fig:short_term_results}c) exhibits a broad, right-skewed distribution extending beyond the periphery peak, indicating larger-scale and strong spatial coherence in the velocity field. Beyond this range, the $c_1(g,L)$ centerline drops below $0.5$ again and the flow transitions toward rougher behavior, marking the onset of far-wake dynamics. Turning now to the lateral characterization, the wake core profiles ($-2<y/D<2$) at $\sim 4D$-$6D$ display a \emph{W-shaped} modulation in both  $c_1(g,L)$ and $c_2(g,L)$, reflecting the well-known double-Gaussian lateral velocity-deficit structure of the wake \cite{qian2025novel}. Moreover, both of the $c_2(g,L)$ profiles (blue dashed lines in Fig.~\ref{fig:short_term_results}f) in this region cross the zero mark between approximately $2D$ and $3D$ distances in the lateral $y/D$ direction, indicating the presence of the inverse-cascading tendency described in the previous subsection. For lateral positions beyond $3D$, the $c_2(g,L)$ values return to negative, consistent with the expected free-flow atmospheric condition.

\subparagraph{3. Far-wake, transition to next zone around $8.5<x/D<10$}
The wake continues to weaken but remains dynamically active, typically showing $0 < c_1(g,L) < 0.5$ in Fig.~\ref{fig:short_term_results}a. Concurrently, the centerline profile of $c_2(g,L)$ reveals an additional intermittency cycle (Fig.~\ref{fig:short_term_results}b), centered at $\sim7$–$8D$ and strongly damped by $\sim9D$. Consistently, the lateral $c_2(g,L)$ profiles (Fig.~\ref{fig:short_term_results}f) exhibit a pronounced dip at $\sim6D$ and retain weak organization up to $\sim8D$, marking the far-downstream extent of this regime. On average, the inverse-cascade signature is apparent near $\sim6D$ and progressively diminishes toward $\sim8D$.
\begin{figure}[!t]
    \makebox[\textwidth][c]{%
        \begin{minipage}{1\linewidth}
            \centering
            \includegraphics[width=\linewidth]{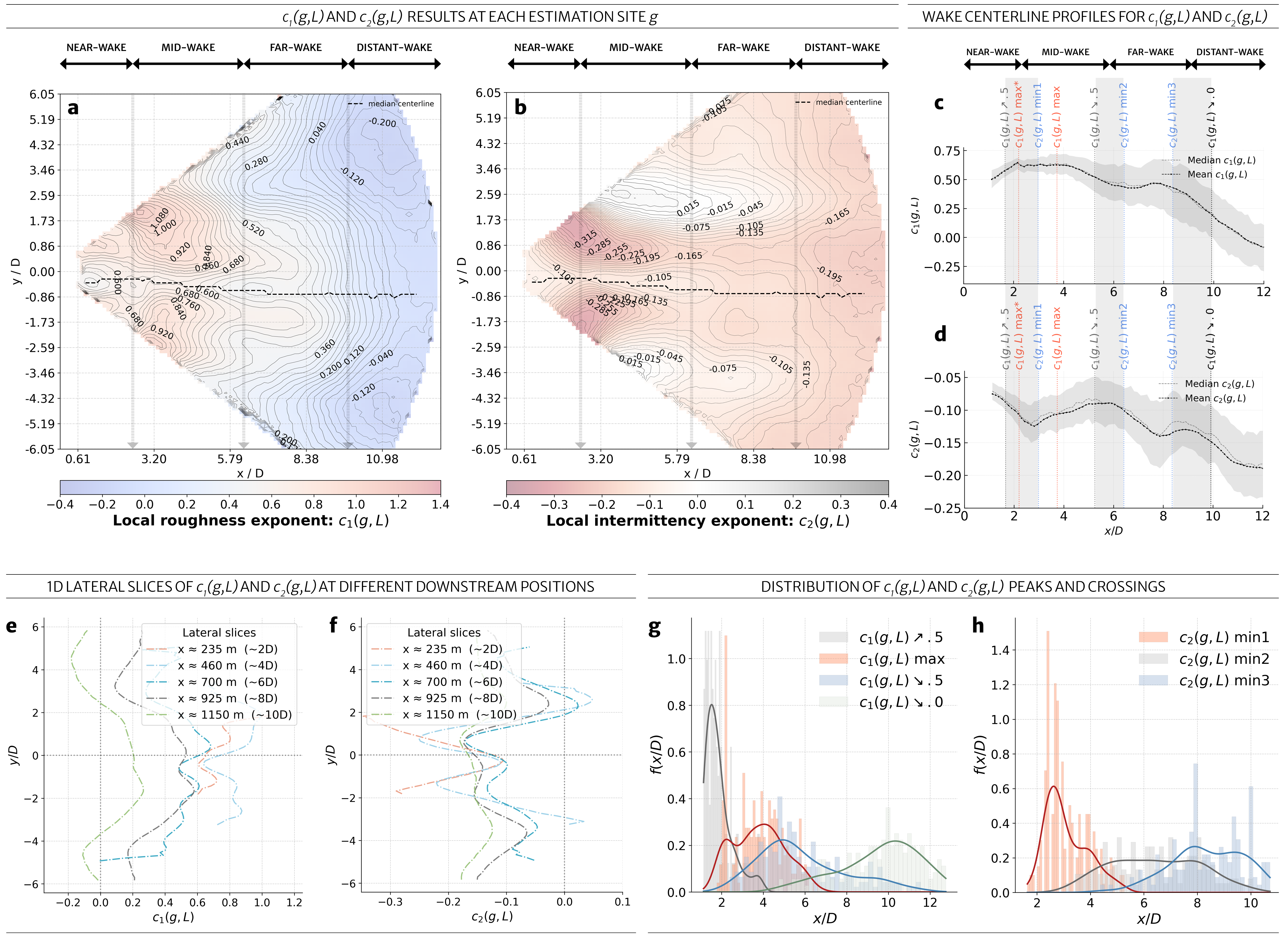}
        \end{minipage}
    }
    \caption{\captiontitle{Long-term results}: Averaged results for $c_{1}(g,L)$ (a,c) and $c_{2}(g,L)$ (b,d) across 222 scans, corresponding to approximately two hours of data with an average inflow wind speed of $6.94~\mathrm{m,s^{-1}}$ (ED Fig.~\ref{fig:mfcup_48}a). The global Pearson correlation between the two fields (a,b) is $\rho = -0.03$, indicating that they capture largely distinct structural properties. For comparison, the correlation between the long-term averaged CWES metrics in ED Fig.~\ref{fig:ext_TI_VD_long} is $\rho = -0.77$. The centerlines in panels (a,b) are obtained by taking the median $y$-coordinate across all individually extracted centerlines. Panels (g,h) show the distribution of landmark points derived from the streamwise profiles in (c,d), all of which jointly contribute to characterizing the wake zones.}
    \label{fig:main_c1c2_long}
\end{figure}
\subparagraph{4. Distant-wake}
With respect to the centerline statistics, $c_1(g,L)$ (Fig.~\ref{fig:short_term_results}a) approaches noise-like values, reflecting very short-ranged correlations. This behavior is consistent with the lateral profiles of $c_1(g,L)$ at $\sim10D$, where the values resemble those expected at the periphery. A similar trend is observed for the lateral profile of $c_2(g,L)$, which has almost entirely lost its W-shaped character. Together, these tendencies indicate an increasing similarity to ambient atmospheric conditions, matching on average both the peak of the histogram in the peripheral regions (Fig.~\ref{fig:short_term_results}c) and the cup-anemometer measurements presented in the following section. In this regime, any additional intermittency cycle is weak or absent, and the statistical contrast between the wake and the ambient flow is correspondingly reduced.

\subparagraph{Comparison to CWES metrics} Projecting the wake zones identified above back onto the instantaneous velocity-deficit and turbulence-intensity maps in Fig.~\ref{fig:main_c1c2}a and Fig.~\ref{fig:main_c1c2}b (grey dashed lines) shows that the above derived segmentation of the wake can indeed be visually motivated from these fields commonly used in CWES, most clearly from the velocity-deficit map. The latter exhibits regions of gradual downstream recovery patterns that appear to form according to the extracted wake zones. However, the transitions between these regions remain diffuse in the classical variables: 
In the short- (Fig. \ref{fig:main_c1c2}c,d) and long-term results (ED Fig.~\ref{fig:ext_TI_VD_long}c,d), the velocity-deficit profile is a single, steadily increasing curve as it is approaching towards free-flow conditions, and the TI profile decreases smoothly, with neither exhibiting distinctive features, crossings, or inflection points that would signal meaningful structural transitions within the wake. As a result, CWES metrics provide only a limited basis for identifying the differentiated wake regions revealed by the joint use of LMF parameters.

\paragraph{Long term results or the persistent structural organization of the wake}
The contrast between the short-term and long-term results serves here to highlight the interplay between transient wake dynamics and persistent structural features. 

\subparagraph{Lateral profiles} Across the lateral profiles at hub height, the short-term  (Fig.~\ref{fig:short_term_results}d and Fig.~\ref{fig:short_term_results}f) and long-term  (Fig.~\ref{fig:main_c1c2_long}e and Fig.~\ref{fig:main_c1c2_long}f) results show remarkable agreement. An important distinction is that the long-term profiles are smoother and nearly symmetric about the centerline, indicating that meandering-induced lateral shifts have been largely averaged out. Notably, even in the long-term analysis, the wake remains detectable far downstream: at $\sim 10D$, the $c_1(g,L)$ (Fig.\ref{fig:main_c1c2_long}e) field still exhibits a coherent wake signature, with elevated $c_1(g,L)$ values within the core wake region rather than reverting fully to ambient conditions. This underscores the importance of the distant-wake zone.

\subparagraph{Streamwise profiles along the wake centerline} The long-term centerline profiles in Fig.~\ref{fig:main_c1c2_long}c and Fig.~\ref{fig:main_c1c2_long}d confirm that the intermittency cycles observed in the 10-scan averages are not long-term features of the wake. In the short-term data, $c_2(g,L)$ shows multiple localized drop-recovery oscillations, whereas in the long-term results the number of these cycles are reduced, leaving only the dominant minimum near $\sim 2D$ and the broad drop-recovery between $\sim 2D$ and $6D$ (Fig.~\ref{fig:main_c1c2_long}d). This indicates that the intermittency cycles stem from more short-lived, localized events and instabilities and disappear under long-term averaging, highlighting the importance of instantaneous analysis for resolving the full wake dynamics. 

\subparagraph{Wake zones} The above tendencies are reflected in the four wake regions in Fig.~\ref{fig:main_c1c2_long}: sharp gradients in the near-wake; a broad plateau of elevated $c_1(g,L)$ together with a persistent drop-recovery pattern in $c_2(g,L)$ between roughly $2D$ and $5D$ in the mid-wake; and a gradual weakening of coherence with a mild decline in $c_2(g,L)$ throughout the far-wake, before the fields become nearly uniform beyond $\sim 10D$, signaling the far-end of the distant-wake. The distributions of the $c_1(g,L)$ and $c_2(g,L)$ landmark positions in Fig.~\ref{fig:main_c1c2_long}g and Fig.~\ref{fig:main_c1c2_long}h provide independent statistical support for this segmentation: the first $c_1(g,L)$ maximum and its adjacent shoulders cluster between $2D$ and $5D$ (Fig.~\ref{fig:main_c1c2_long}g), whilst the downstream $c_1(g,L)$ zero-crossings form a broad peak near $10D$, consistent with the far-wake-to-distant-wake transition. Likewise, the first intermittency minimum of $c_2(g,L)$ is concentrated around $2D$, marking the near- to mid-wake transition, while the second and third minima predominantly occur around $6D$ and $10D$, aligning with the far-wake region where intermittency progressively weakens and eventually vanishes. Notably, the histogram of the second minimum (grey color in Fig.~\ref{fig:main_c1c2_long}h) is extremely broad, highlighting the diversity and variability of meandering-driven dynamics in this part of the wake. Overall, the agreement between the two aggregation levels - visible across both the two-dimensional fields (ED Fig. \ref{fig:ext_c1c2_shortterm} and Fig. \ref{fig:main_c1c2_long}) and the one-dimensional centerline and lateral profiles (Fig. \ref{fig:short_term_results} and Fig. \ref{fig:main_c1c2_long}) - provides robust confirmation of the four-zone structure inferred from the short-term analysis.

\subparagraph{Second case study} A second case study, shown in ED Fig.~\ref{fig:ext_second_cs}, was chosen on the same day during a measurement period between 03:00 and 05:00 local time. At the time of the instantaneous scan (04:23; ED Fig.~\ref{fig:ext_second_cs}a-d), the inflow wind speed was $7.62~\mathrm{m\,s^{-1}}$, compared with $6.19~\mathrm{m\,s^{-1}}$ in Fig.~\ref{fig:main_c1c2}. Consistent with this difference, the near wake is slightly shortened by approximately 10$\%$, and elevated roughness levels - remaining within the previously defined $0.0 < c_1(g,L) < 0.5$ - persist in the far-wake region (ED Fig.~\ref{fig:ext_second_cs}c), while three intermittency cycles are forming (ED Fig.~\ref{fig:ext_second_cs}d), similar to the main case study. This elevated $c_1(g,L)$ at higher wind speeds in the far-wake zone is in line with a Davenport-like coherence model, which predicts a reduced rate of exponential decay at higher wind speeds~\cite{thedin2022investigations}. Despite these differences in roughness levels, the wake-zone landmarks (ED Fig.~\ref{fig:ext_second_cs}g,h,k,l) remain remarkably robust across the two case studies when considering the long-term results.

\begin{figure}[htbp]
    \makebox[\textwidth][c]{%
        \begin{minipage}{1.\linewidth}
            \centering
            \includegraphics[width=\linewidth]{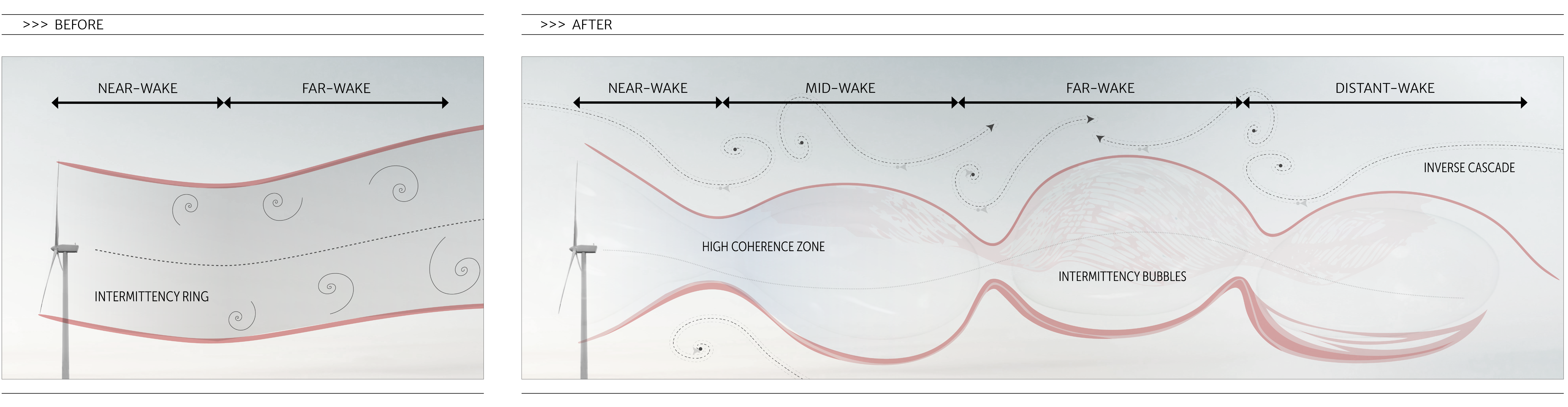}
        \end{minipage}
    }
    \caption{\captiontitle{Schematic overview of the derived wake morphology and dynamics}. The image on the left illustrates the previously assumed ``intermittency ring'' co-evolving with the wake centerline (dashed line), whereas the image on the right provides an overview of the newly derived wake dynamics, referred to as ``intermittency bubbles''. The four distinct wake zones, the high-coherence mid-wake region, and the presence of inverse-cascade processes are also key features of the new morphology.}

    \label{fig:visual}
\end{figure}
\paragraph{Synthesis: toward a new morphology of wake structures}

The joint use of the roughness exponent $c_1(g,L)$ and the intermittency exponent $c_2(g,L)$ reveals spatial wake-flow dynamics that are not captured by CWES metrics, and is therefore essential for robustly identifying the local spatial organization of wake structures. We identified four new findings concerning the wake morphology which we summarized in Fig.~\ref{fig:visual}: (1) Whereas previous research~\cite{vermeer2003wind} has primarily focused on only two regions - the near- and far-wake - and occasionally their transition zone, the multifractal characterization presented here reveals four clearly distinct wake zones together with their respective transition regions. (2) Moreover, the $c_1(g,L)$ field uncovers a pronounced mid-wake zone of high spatial coherence, which may have important implications for turbine spacing and wind-farm layout design. (3) The intermittency field $c_2(g,L)$ further shows that localized turbulence activity is not organized as a simple intermittency ring running approximately parallel to the wake centerline~\cite{neunaber2020distinct,bodini2017three}, but instead forms localized “bubbles’’ of intermittency that do not reside exclusively along the wake/free-flow interface; rather, they periodically intrude into the wake interior at several downstream positions. These structures were earlier termed “cycles’’ in this work during the process of establishing evidence for their behaviour. (4) Finally, the emergence of a pronounced spatial region of vanishing or even positive $c_2(g,L)$ at the wake boundary indicates an inverse-cascade-like behavior, which acts as a buffer layer mediating the transition between the highly intermittent wake boundary and the surrounding ambient flow. Collectively, these features may greatly influence fundamental wake dynamics such as wake meandering and recovery.

\section*{Building robustness through comparison with classical multifractal and turbulence analysis of cup anemometer data}

Here, we compare the results from the previous section with classical, well-established methodologies based on global analyses of temporally resolved one-dimensional datasets. The aims are \emph{i.)} to determine whether the roughness of the peripheral LiDAR results matches the roughness levels obtained here; \emph{ii.)} to further understand the coexistence of low roughness and pronounced intermittent behavior; and \emph{iii.)} to evaluate whether the classical inertial range, if it exists, represents the dominant scaling regime for robust feature extraction.

\begin{figure}[h]
    \makebox[\textwidth][c]{%
        \begin{minipage}{.75\linewidth}
            \centering
            \includegraphics[width=\linewidth]{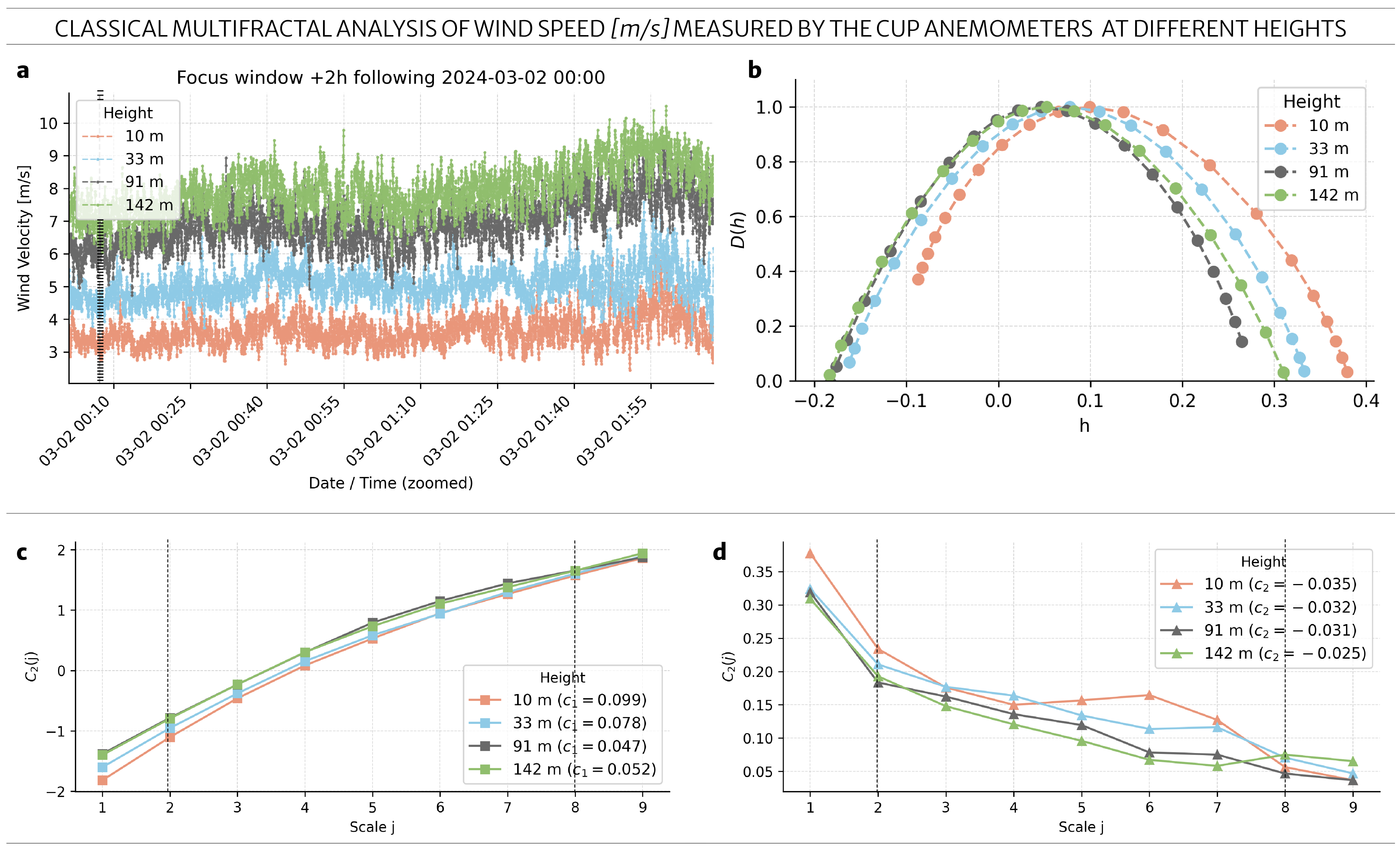}
        \end{minipage}
    }
    \caption{\captiontitle{Classical multifractal analysis} Wavelet-leader multifractal spectra (b) computed from the cup-anemometer velocity signals (a) within the $2$h window of the scan time (black dashed line) shown in Fig.~\ref{fig:main_c1c2}. Panels (c) and (d) show the corresponding scaling of $C_1(j)$ and $C_2(j)$ (see Methods), respectively, along with the estimated $c_1$ and $c_2$ exponents obtained using the scale range indicated by the grey dashed lines. The corresponding analysis for the $48$h window is displayed in ED Fig. \ref{fig:mfcup_48}.}
    \label{fig:mfcup_24}
\end{figure}

\paragraph{Classical multifractal and turbulence analysis}
In Fig.~\ref{fig:mfcup_24} (and ED Fig.~\ref{fig:mfcup_48}), we therefore apply classical multifractal analysis (see the corresponding sections in Methods) to cup-anemometer wind-velocity data measured at four heights and at two aggregation levels: $2\mathrm{h}$, corresponding to the long-term results shown in Fig.~\ref{fig:main_c1c2_long}, and $48\mathrm{h}$, corresponding to the two-day period around the scan time shown in Fig.~\ref{fig:main_c1c2}. For the $48\mathrm{h}$ aggregation level, we additionally conducted classical turbulence analysis, the results of which are shown in ED Fig.~\ref{fig:ext_turb}. The resulting multifractal spectra in Fig~\ref{fig:mfcup_24}b reveal a height dependence in both width and peak location: lower levels exhibit broader, more asymmetric (at 10m) spectra with peaks at larger $h$ values, indicative of smoother dominant fluctuations and stronger intermittency, whereas higher levels display peaks at smaller $h$, reflecting comparatively rougher dominant structures. These patterns appear in both the $48\mathrm{h}$ (Fig.~\ref{fig:mfcup_48}b) and the $2\mathrm{h}$ focus-window analyses.

The derived multifractal spectra also show that fluctuations are simultaneously very rough (as indicated by the peak location at both aggregation levels) and distinctly non-Gaussian (as reflected by the spectral width). This is consistent with the LiDAR results in Fig.~\ref{fig:short_term_results}c, where the periphery region (orange) exhibits a clear histogram peak of $c_{1}(g,L)$ between $0.0$ and $0.1$, closely matching the spectral peaks in Fig.~\ref{fig:mfcup_24}. Similar roughness exponents have been reported by Cadenas et al.~\cite{cadenas2019wind}, who observed long-term wind-speed variability exhibiting comparably low Hurst-like exponents. In both data sets used here (cup and LiDAR), non-Gaussianity persists even though the extracted $c_1(g,L)$ values are far smaller than those expected under classical Kolmogorov K41 phenomenology~\cite{Frisch1995}, which suggests a Hurst exponent of $H=1/3$, demonstrating that heavy-tailed, intermittent processes remain active despite comparatively weak second-order scaling.
\paragraph{Roughness and intermittency beyond the inertial range}
It must be stressed here that the fitting range used for the LiDAR analysis ($0.47$–$1.47D$) is substantially larger than the classical inertial range of turbulence, defined by the scale-invariant Kolmogorov regime between energy injection and viscous dissipation scales~\cite{Frisch1995,kolmogorov1991local}. In the atmospheric surface layer, this inertial range - if it exists at all - is typically confined to a narrow band of sub-second to few-second scales~\cite{KaimalFinnigan1994,Wyngaard2010,Stull1988}, making it difficult to resolve with finite-resolution measurements. This limitation is evident in both the LiDAR and, to a lesser extent, the cup-anemometer data: although ED Fig.~\ref{fig:ext_turb}a shows a brief Kolmogorov-like scaling at the smallest resolved lags, the continuously varying local slope in ED Fig.~\ref{fig:ext_turb}b indicates that no well-defined inertial range forms. Accordingly, the dominant scaling observed in our data (green dashed line in ED Fig.~\ref{fig:ext_turb}a) lies well beyond this range, spanning approximately $8$ to $600$ s, and reflects a regime rougher than predicted by idealized turbulence theory. Real surface-layer flows rarely satisfy homogeneity or isotropy; instead, shear, buoyancy, and mesoscale processes inject energy across a broad range of scales, further inhibiting the formation of a clean inertial range~\cite{Wyngaard2010,Stull1988}. This complexity motivates the use of wavelet-based methods, which do not rely on strong \textit{a priori} assumptions on the increment statistics~\cite{abry2003scaling,argoul1989wavelet}. Importantly, although velocity fluctuations remain rough across a wide range of scales, yielding low Hurst exponents, this does not preclude multifractal transport: the clear scaling of $C_1$ and $C_2$ (eqs.~\ref{eq:C1} and~\ref{eq:C2}; Fig.~\ref{fig:mfcup_24}, ED Fig.~\ref{fig:mfcup_48}, and ED Fig.~\ref{fig:ext_method}b,c) demonstrates multifractality across all resolved scales, including those beyond the classical inertial range, consistent with previous studies~\cite{Muzy2010,LovejoySchertzer2013}.

\section*{Discussion}

\subparagraph{Methodology}
Concerning the methodological implementation, this study demonstrates that localized multifractal analysis can be applied directly to non-homogeneously sampled instantaneous LiDAR scans to extract meaningful information that complements classical wind energy science metrics. This capability is particularly valuable for operational wind-energy applications, where data resolution is often limited, clean inertial ranges are typically absent, and spatiotemporal sampling may be irregular. Moreover, the strong agreement between LiDAR-based periphery statistics and the multifractal spectra of one-dimensional cup-anemometer signals further highlights the robustness of this spatial approach.

\subparagraph{Wake structure}
A key conceptual outcome is that the turbine wake is far more structured than suggested by velocity deficit or turbulence intensity alone. The joint use of $c_1(g,L)$ and $c_2(g,L)$ reveals four distinct wake regions - near-, mid-, far-, and distant-wake - with non-trivial transitions between them. The mid-wake, in particular, emerges as a coherent zone with long-range spatial correlations, while intermittency appears in localized “bubbles’’ that periodically intensify and collapse within the wake interior. These features refine the classical notion of an “intermittency ring’’ and point to dynamically active processes that remain hidden to traditional, strongly averaged metrics and to standard turbulence analysis based on velocity increments. The appearance of a broad zone of vanishing or positive $c_2(g,L)$ at the wake boundary further suggests an inverse-cascade-like behaviour that mediates the interaction between the wake and the ambient atmosphere.  

\subparagraph{Limitations}
A primary limitation of this study is that the analysis was restricted to stable nocturnal conditions and a single inflow sector; extending it to convective regimes~\cite{bodini2017three}, complex terrain~\cite{barthelmie2019automated}, or yawed inflow~\cite{herges2017high} would further reveal the distinct traits and drivers of the observed zonal structure. Further, on the one hand, the available resolution appears insufficient to clearly isolate an inertial subrange, as evidenced by generally reduced roughness exponents across the datasets. This suggests that higher-resolution spatial measurements would be valuable in future work. Along similar lines, the present nacelle-based LiDAR setup is restricted to the radial component of the wake flow, whereas multi-LiDAR field measurements \cite{iungo2013field,wildmann2018wind} would enable resolution of the full wind-velocity vector within the turbine wake. On the other hand, the behaviour observed here highlights the strengths of wavelet-based multifractal analysis in extracting scaling exponents and intermittency from nonstationary, noisy, or mixed-regime atmospheric measurements. 

\subparagraph{Outlook and industry relevance}From an industry perspective, these findings may provide complementary information to existing wake diagnostics. The strongly coherent mid-wake region and the inferred spatial distribution of intermittency could inform turbine spacing, wake-steering strategies, and loading assessments. Because the framework operates with high computational efficiency on instantaneous PPI scans, it is, in principle, compatible with real-time monitoring and control. Looking ahead, these results suggest that wake models commonly used in load-estimation pipelines may need to be augmented with additional parameters to capture the pronounced spatial non-homogeneity identified here. Finally, the baseline established for local roughness and intermittency analysis may provide a solid foundation for future studies investigating how these features influence key wake processes, including wake meandering and recovery.

{\footnotesize
\bibliographystyle{unsrt}
\bibliography{sn-bibliography}
}

\clearpage

\newpage
\section*{\fontsize{14}{18}\selectfont Methods}
\setcounter{equation}{0}
\renewcommand{\theequation}{M\arabic{equation}}
The following section presents the Methods of the associated main article, ``The spatial organization of turbine wakes.'' We first provide a detailed description of the data and recapitulate the main methodology in greater depth. We then introduce the classical multifractal and turbulence analyses applied to one-dimensional wind-velocity measurements recorded by cup anemometers. Finally, we describe the algorithm used to detect the wake centerline.

\subsection*{Data description and atmospheric conditions}
This study relies on two complementary data sources, both obtained from the WiValdi wind energy research site in Germany \cite{DLR_WiValdi}. The first source is a Doppler wind LiDAR installed on the nacelle of a wind turbine, while the second is a meteorological mast equipped with cup anemometers at multiple elevations and a wind vanes positioned near the turbine hub height.

\paragraph{Nacelle-mounted LiDAR}
The dataset was acquired using a Leosphere Windcube 200S nacelle-mounted Doppler wind LiDAR operated by the German Aerospace Center (Deutsches Zentrum für Luft- und Raumfahrt, DLR). The device was mounted on the nacelle of an Enercon E-115 wind turbine with rotor diameter $D=115.7$ m at the WiValdi site, at an altitude of $96$ m above sea level. In this setup, the LiDAR was oriented downstream to measure the wind turbine wake. With respect to date and time scale, our primary example is a single instantaneous scan obtained on 2 March 2024 at 00:07. For a more robust statistical analysis, however, we consider 222 scans recorded between 00:00 and 02:00 on the same day. Details of the instrument settings used during the measurements are provided in a separate study (J. Menken $\&$ N. Wildmann, in preparation). A single scan takes approximately 5.83 s, while the physical pulse length of the LiDAR is 50 m. Consequently, each line-of-sight velocity estimate within a single range gate represents a weighted average over the LiDAR pulse length. This limits the smallest spatial scales that can be resolved in the radial direction and was taken into account when selecting the minimum analysis length scale $r_{min}$ (see Main Methodology below). Local outlier detection was performed using a wavelet-like quantity, $\kappa_e-\mu(e,r_{min})$. Common challenges when using nacelle-mounted LiDAR data include accounting for the motion of both the turbine and the LiDAR system during measurements, as well as dealing with noisy or incomplete data. In addition, hard targets, such as meteorological masts or the neighboring wind turbine within the wind park can strongly contaminate the measurements. These issues are discussed in the following paragraph.

\paragraph{Meteorological mast: cup anemometers and wind vane data}
To support and benchmark our nacelle-mounted LiDAR results against the undisturbed inflow turbulence, we additionally use data from cup anemometers sampled at 1~Hz and mounted on a meteorological (met) mast located southwest of the turbine at a distance of approximately 2D. For an inflow wind direction of $124-157^\circ$, this geometry ensures that the selected LiDAR scans are not disturbed by the mast. Cup anemometer measurements are available at heights of $10$~m, $33$~m, $91$~m, and $142$~m. As shown in ED Fig.~\ref{fig:ext_cup_alpha}, we analyzed met mast data collected from 1 March 2024 at 00:00 to 3 March 2024 at 00:00, so that the time window used for the nacelle LiDAR analysis lies approximately at the center of the measurement period. Wind vane records from the same 48-hour interval provide wind direction measurements at height $88$~m (ED Fig.~\ref{fig:ext_cup_alpha}).

\paragraph{Atmospheric conditions at time of analysis}
On 2 March 2024, between 00:00 and 02:00 local time, the inflow winds were southeasterly ($124$--$157^\circ$; ED Fig.~\ref{fig:ext_cup_alpha}a). During this period, wind speeds at hub height ranged from $4.75$ to $9.45~\mathrm{m\,s^{-1}}$, as measured by the cup anemometer at $91~\mathrm{m}$ (Fig.~\ref{fig:mfcup_24}), with a mean wind speed of $6.94~\mathrm{m\,s^{-1}}$. The vertical wind-speed profile exhibited shear exponents ranging between $0.27-0.33$ (ED Fig.~\ref{fig:ext_cup_alpha}b, grey focus window), derived from the four cup anemometer signals (ED Fig.~\ref{fig:mfcup_48}a), which is consistent with stable nocturnal boundary-layer conditions. As, to our knowledge, this is the first analysis to extract localized higher-order structural parameters from non-homogeneously spaced LiDAR samples, a stable atmospheric regime provides an adequate starting point, as it helps exclude additional sources of complexity such as rapid wind-direction shifts (veer) or transient shear events.
\subsection*{Main methodology}
We differentiate between two methodological components: (1) one that provides a spatially localized summary of commonly used metrics in classical wind energy science (CWES), and (2) another that derives the local multifractal (LMF) properties using a recently developed methodology designed to robustly handle non-homogeneous spatial point processes \cite{Lengyel2022_localMultifractalityUrban, lengyel2023roughness,lengyel2025bivariate}. In both methodologies, we define the LiDAR line-of-sight velocity data $\kappa_e$ as a point process $e(x_e, y_e)$, characterized by the spatial coordinates $x_e$ and $y_e$, and an associated value $\kappa_e$ which in this case is the radial component of the wind field [$\mathrm{m\,s^{-1}}$] measured by the LiDAR. Over the entire area to be analyzed, we overlay an equidistant grid $g(x_g, y_g)$, which serves as the spatial framework for computing all CWES and LMF parameters. To maximize correspondence between the two methodologies, we compute localized quantities in both approaches and therefore define a weighting function between the original point process $e$ and the equidistant grid $g$ as $w_{g,e} = 1$ if $d_{g,e} < L$, and $0$ otherwise, where $d_{g,e} = \sqrt{(x_e - x_g)^2 + (y_e - y_g)^2}$, and then normalizing such that $\sum_e w_{g,e} = 1$. ED Fig.~\ref{fig:ext_method} summarizes the methodology and its main components.

\paragraph{Wind energy science} The characteristics in context of classical wind energy science computed here are the localized mean and standard deviation of the radial wind speed, along with the velocity deficit and turbulence intensity. These quantities are defined in the main manuscript in eq. \ref{eqn:I}, eq. \ref{eqn:sigma}, eq. \ref{eqn:VD}, and eq. \ref{eqn:TI}. For the velocity-deficit field in eq. (\ref{eqn:VD}), we estimate the inflow wind speed using the maximum of the local means $\mu(g, L)$ within each instantaneous scan. This assumption is supported by the cup-anemometer data shown in Fig.~\ref{fig:mfcup_48} and provides a convenient and robust basis for processing a large number of scans using only a single data source. The resulting two-dimensional velocity deficit is then used to extract the wake centerline, which we subsequently overlay on all other derived features. A detailed description of the computational procedure used to extract the centerline is provided in the last subsection of Methods. We also consider turbulence intensity ($TI(g, L)$ in eq. (\ref{eqn:TI})) for each PPI scan, which in CWES aims to quantify wind-speed variability via relating the standard deviation of the velocity to its mean. It characterizes the unsteadiness of the flow and serves as a widely used indicator of turbine loading, fatigue, and also wake behaviour \cite{Burton2011, IEC61400-1_2019}. 

\paragraph{Local multifractal analysis}
The local multifractal analysis framework \cite{lengyel2023roughness,lengyel2025bivariate} was specifically developed to extend classical multifractal methods (described below in Methods) to address the challenges of non-homogeneously sampled point processes. It consists of two closely connected steps. The initial processing step relies on a transformation that handles non-uniform spatial sampling and follows principles analogous to those of wavelet decomposition. This stage builds on earlier studies \cite{Lengyel2022_localMultifractalityUrban,lengyel2023eusipco} that demonstrated the robustness of LMF parameter estimation for irregular point patterns and clarified the distinction between processes defined solely on the spatial distribution or support of the points and those that also incorporate marks, i.e., values attached to the point locations. The second step characterizes the behavior of data subsets across local neighborhoods - here, portions of the PPI scan – to infer the local dynamics of the wake. 

\subparagraph{Wavelet decomposition} The main ingredients of the local multifractal analysis are multiresolution quantities centered at each original point $e$ and computed across successive length scales $r$. The number of observations within $r$ is defined as $N(e, r) = \operatorname{card}\{ e' : d_{e,e'} \leq r \}$, where $d_{e,e'} = \sqrt{(x_e - x_{e'})^2 + (y_e - y_{e'})^2}$ denotes the Euclidean distance, and $\operatorname{card}$ indicates the number of observations. The mean radial wind speed at scale $r$ centered at $e$ is denoted $\mu(e,r)$ (Eq.~\ref{eqn:sb_mark}), with the corresponding two-dimensional flat-hat wavelet-like quantity given by $D(e,r)$ (Eq.~\ref{eqn:wt_hat_mark}), as described in the associated main manuscript.

It must be noted that if $N(e)$ is the total number of points $e$ then $\frac{1}{N(e)}\sum_e D(e,r) =0$. In order to robustly assess complex spatial dependencies that go beyond second-order statistics, more elaborate multiscale quantities need to be deployed, referred to as wavelet p-leaders \cite{wendt2007bootstrap,wendt2009wavelet, leonarduzzi2016p,jaffard2016p}:  they make it possible to quantify the sets where the data attain a specific pointwise regularity exponent. For non-homogeneous two-dimensional point processes, with $p=2$, these quantities have been recast \cite{lengyel2025bivariate} as $T(e,r,p)$ in accordance with eq. \ref{eqn:leader} in the main manuscript.

\subparagraph{Scaling functions} Moving on, to observe the local scaling characteristics, this framework calculates the first and second order cumulants of the logarithm of the wavelets and wavelet leaders as described in eq. \ref{eq:C1} and \ref{eq:C2} in the main manuscript and repeated here for clarity;
\begin{align}
    C_1(g,r,L) &= \sum_{e} w_{g,e} \, \log |D(e,r)| 
    \sim c_1(g,L) \, \log(r) + k_1(g,L),
    \label{eq:m_C1} \\
    C_2(g,r,L) &= \sum_{e} w_{g,e} 
    \left[\log |T(e,r,p)| - C_1(g,r,L)\right]^2 
    \sim_{r\to0} c_2(g,L) \, \log(r) + k_2(g,L).
    \label{eq:m_C2}
\end{align}
The exponents $c_1(g,L)$ (using the wavelet $D(e,r)$) and $c_2(g,L)$ (using the wavelet p-leader $T(e,r,p)$) are fitted along the available length scales $r$ at each grid location $g$. This step is crucial, as it ensures that the entire domain is systematically scanned using the local environment $L$, allowing us to extract the exponents consistently across all locations and produce a new feature map composed entirely of these dimensionless quantities \cite{Lengyel2022_localMultifractalityUrban,lengyel2025bivariate}. The first exponent, $c_1(g, L)$, represents the fractal component of the process and is closely linked to the Hurst exponent $H(g, L)$ through the relation $c_1(g, L) = H(g, L) - c_2(g, L)$. Accordingly, it reflects the range of spatial or temporal dependencies in the data. In classical multifractal analysis, $c_1(g,L)$ also marks the peak of the multifractal spectrum (see also Fig.~\ref{fig:mfcup_48}), corresponding to the most commonly occurring local regularity. The second exponent, $c_2(g, L)$, is the multifractal component and represents the width of the multifractal spectrum. It can be interpreted as the intermittency of the data, indicating - within the classical framework - the increasing probability of extreme fluctuations as length scales decrease. Therefore, the joint analysis of $c_1(g, L)$ and $c_2(g, L)$ can be understood as a local summary of the multifractal spectrum at each grid point $g$, captured through its height and width \cite{jaffard2006wavelet,Wendt2007}.  

\newpage
\subsection*{Classical multifractal formalism}
\label{app:mf}
\subsubsection*{Wavelet transform}
Let $u(t)$ denote the velocity signal measured by the cup anemometers with sampling rate $1\,\mathrm{Hz}$, and let $\psi$ be a mother wavelet with $N_\psi$ vanishing moments and fulfilling the admissibility condition $\int_{\mathbb{R}} \psi(t)\, dt = 0$. In this work, we use the Daubechies-3 wavelet, which combines three vanishing moments and compact support in classical regularly sampled datasets, thereby helping to reduce asymmetry and boundary effects. The associated dyadic family of discrete wavelets is
\begin{equation}
    \psi_{j,k}(t)
    = 2^{j/2}\, \psi(2^{j}t - k),
    \label{eq:wavelet_def}
\end{equation}
where $j$ denotes the scale and $k$ the translation. The discrete wavelet transform coefficients are obtained by projection,
\begin{equation}
    d_{j,k}
    = \langle u, \psi_{j,k} \rangle
    = \int_{\mathbb{R}} u(t)\, \psi_{j,k}(t)\, dt.
    \label{eq:wavelet_coeffs}
\end{equation}

\subsubsection*{Wavelet $p$-leaders}

Wavelet p-leaders provide a robust way to characterise local irregularity and multifractality \cite{wendt2007bootstrap,wendt2009wavelet, leonarduzzi2016p,jaffard2016p}, and are defined as
\begin{equation}
    \ell^{p}_{j,k}
    =
    \left(
        \sum_{\substack{j' \le j \\ \lambda' \subset 3\lambda}}
        |d_{j',k}|^{p}
    \right)^{1/p},
    \label{eq:pleader_def}
\end{equation}
where $3\lambda$ denotes the union of $\lambda$ and its two neighbouring dyadic intervals, with $\lambda_{j,k} = [k2^j, (k+1)2^j] $. We use $p=2$ in this work.

\subsubsection*{Scaling functions and cumulants}

For each scale $j$, the wavelet-leader structure functions are defined as
\begin{equation}
    S_\ell(q,j)
    =
    \frac{1}{n_j} \sum_{k} |\ell^{p}_{j,k}|^{q},
    \label{eq:Sqj}
\end{equation}
where $n_j$ is the number of multiresolution quantities (wavelet p-leaders) at that scale. Scale invariance
implies the power-law behaviour
\begin{equation}
    S_\ell(q,j)
    \sim 2^{\,j\,\zeta(q)},
    \label{eq:zeta_def}
\end{equation}
which defines the scaling function $\zeta(q)$. The cumulants of order $m$ of $\log \ell_{j,k}^p$ follow \cite{wendt2007bootstrap,wendt2009wavelet}
\begin{equation}
    C_m(j)
    =
    \mathrm{cum}_m(\log \ell^p_{j,k})
    \sim
    c_m\, j + \mathrm{k_m},
    \label{eq:cumulants}
\end{equation}
where $c_1$ corresponds to the dominant local regularity and $c_2$
quantifies intermittency.

\subsubsection*{Multifractal spectrum}
The Legendre spectrum in one dimension is defined as
\begin{equation}
    D(h)
    =
    \min_{q \ne 0}
    \bigl(
        1 + q\,h - \zeta(q)
    \bigr).
    \label{eq:wll_legendre}
\end{equation}
Under standard regularity assumptions, this Legendre spectrum forms an upper bound for the true Hölder-based multifractal spectrum and, within the wavelet-leader multifractal formalism \cite{Wendt2007} it is used as its numerical estimator.

\subsubsection*{Practical estimation of the wavelet-leader multifractal spectrum}

In this work, the multifractal spectrum is estimated directly \cite{jaffard2006wavelet,Wendt2007} without a numerical Legendre transform \cite{Wendt2007}. For each moment of order $q$,  $p_{j,k}(q)$ is defined as
\begin{equation}
p_{j,k}(q)
=
\frac{|\ell^p_{j,k}|^q}{\sum_{k} |\ell^p_{j,k}|^q}.
\end{equation}
The two intermediate scale-dependent quantities are then constructed:
\begin{equation}
V(j,q)
=
\sum_k p_{j,k}(q)\,\log_2 |\ell^p_{j,k}|,
\end{equation}
\begin{equation}
U(j,q)
=
\log_2 n_j + \sum_k p_{j,k}(q)\,\log_2 p_{j,k}(q).
\end{equation}
Under standard scaling assumptions, these quantities behave linearly with $j$:
\begin{equation}
V(j,q) \sim h(q)\,j,
\qquad
U(j,q) \sim (D(q)-1)\,j.
\end{equation}
The Hölder exponents $h(q)$ and the fractal dimensions $D(q)$ are obtained as the slopes of these linear regressions. The multifractal spectrum is finally derived in a parametric form as
$
\mathcal{L}(h) = \{(h(q),D(q)) : q \in \mathbb{R}\}
$, providing a compact summary of the dominant regularity and intermittency of $u(t)$. It must also be noted that under the log-cumulant expansion of the multifractal spectrum, a quadratic approximation of the scaling function,
$\zeta(q)\approx c_1 q + \tfrac{c_2}{2}q^2$, results in a parabolic approximation of the spectrum whose
maximum is located at $h=c_1$ and whose width is governed by $c_2$.

\subsection*{Classical turbulence analysis}
Similar to the classical multifractal analysis section above, in ED Fig.~\ref{fig:ext_turb} we analyze the velocity signal \(u(t)\) measured by a cup anemometer at a sampling frequency \(f_s = 1\,\mathrm{Hz}\) at height of \(91\,\mathrm{m}\). 

The second-order structure function is estimated directly from the discrete signal as
\begin{equation}
S_2(\tau)
    = \frac{1}{N-\tau} \sum_{t=1}^{N-\tau}
      \big(u(t+\tau) - u(t)\big)^2,
    \qquad
    \tau = 1,\dots,\tau_{\max}.
\end{equation}
Assuming power–law scaling within the inertial subrange,

\begin{equation}
\log S_2(\tau) = \zeta_2 \log \tau + k_2.
\end{equation}
the scaling exponent \(\zeta_2\) is obtained from a linear regression of
\(\ln S_2(\tau)\) versus \(\ln \tau\). The associated Hurst-like exponent is $H = {\zeta_2}/{2}$. To capture the scale dependence of the statistics, we also compute the local slope
\begin{equation}
\zeta_2(\tau) = \frac{\mathrm{d}\ln S_2(\tau)}{\mathrm{d}\ln\tau},
\qquad
H(\tau) = \frac{\zeta_2(\tau)}{2},
\end{equation}
using a numerical derivative of \(\ln S_2(\tau)\) with respect to \(\ln\tau\). Additionally, the (normalized) autocorrelation function of the globally detrended velocity fluctuations \(u'(t) = u(t)-\overline{u}\) is computed as
\begin{equation}
\rho(\tau) =
\frac{\langle u'(t)\,u'(t+\tau)\rangle}
     {\langle {u'(t)}^{2}\rangle},
\end{equation}
where the normalization by the variance \(\langle {u'(t)}^{2}\rangle\) ensures that \(\rho(0)=1\) and makes the correlation values directly comparable across time lags. The quantity is evaluated in discrete form for lags up to \(\tau_{\max}=600\) samples. With a sampling frequency of \(1\,\mathrm{Hz}\), this limits the computed structure and autocorrelation functions to $10$ minutes which is in accordance with industrial standards of analysis range.

\subsection*{Shortest-path centerline through a 2D velocity-deficit field.}
\label{sec:app_centerline}
The centerline is extracted using a dynamic-programming shortest-path method, a standard technique for finding minimal energy curves in images \cite{felzenszwalb2012distance}. According to the main article, the measurement plane is discretized into grid points $g = (x_g, y_g)$, and to each grid point we assign the velocity-deficit value VD(g,L) obtained via eq. (\ref{eqn:VD}). To extract the centerline of each scan, we represent the path as a sequence of points
\begin{equation}
(x_0, y_0), (x_1, y_1), \dots, (x_{N_x-1}, y_{N_x-1}),
\end{equation}
where $N_x$ is the total number of streamwise columns. The horizontal coordinate is fixed by the column index:\\ $ x_j \;\text{is the known $x$-coordinate of column }j$
while the vertical coordinate $y_j$ has to be chosen by the algorithm. Each centerline point will correspond to the underlying grid point $ g_j = (x_j, y_j).$
We seek the path minimizing
\begin{equation}
E(\{y_j\})
=
\sum_{j=0}^{N_x-1}
    VD\!\left((x_j,y_j),\,L\right)
\;+\;
\lambda
\sum_{j=1}^{N_x-1} 
    (y_j - y_{j-1})^2 ,
\label{eq:centerline_energy}
\end{equation}
subject to the step constraint
\begin{equation}
|y_j - y_{j-1}| \le M,
\qquad 
j = 1,\dots,N_x-1 .
\label{eq:centerline_constraint}
\end{equation}
Here $\lambda$ is a smoothness parameter that penalizes curvature, and $M$ is the maximum allowed vertical step. In practice, we chose a smoothness parameter $\lambda = 0.0025$ and $M = 3$, which provides sufficient stabilization while still allowing the centerline to follow the underlying velocity–deficit variations closely. To minimize \eqref{eq:centerline_energy} we use dynamic programming. Let $F_j(m)$ be the minimum energy of any partial path ending in row $m$ of column $j$, where $m$ is the discrete row index within a column. The initialization is
\begin{equation}
F_0(m)
=
VD\!\left((x_0,m),L\right),
\end{equation}
and for $j \ge 1$ the recursion is
\begin{equation}
F_j(m)
=
VD\!\left((x_j,m),L\right)
+
\min_{\substack{m' \\ |m-m'|\le M}}
\Bigl[
    F_{j-1}(m') + \lambda (m-m')^2
\Bigr].
\end{equation}
The optimal centerline is obtained by backtracking from $ y_{N_x-1} = \arg\min_m F_{N_x-1}(m)$, which yields all $(x_j, y_j)$ and thus the entire centerline. These coordinates are then overlaid on all extracted features of each processed LiDAR scan to obtain the streamwise centerline profiles along $g_j(x_j, y_j)$ and compute their statistics.

\paragraph{Acknowledgments}
J.L., N.W., J.F., and J.M. acknowledge funding from the German Federal Ministry for Economic Affairs and Energy through the project NearWake (03EE3097C). J.L. further acknowledges support from the DFWIND2 project (03EE3031B), likewise funded by the German Federal Ministry for Economic Affairs and Energy. 

\paragraph{Author contributions}
J.L. performed the computations, prepared the figures and drafted the core manuscript. J.F. assisted in the interpretation of the results from a fluid-mechanical perspective and contributed to manuscript editing. J.L., S.R., P.A., and O.B. contributed to the development of the local multifractal analysis methodology. J.L. and S.R. developed the original computational framework for the local multifractal analysis. P.A. and S.R. established the theoretical and computational foundations of the classical multifractal formalism and the associated wavelet techniques. J.M. and N.W. acquired and curated the LiDAR data and provided technical expertise on the measurement setup. All authors reviewed and approved the submitted manuscript.
\paragraph{Code and data availablity}
The Python code used for the local multifractal analysis is freely available as the Python package LomPy \cite{Lengyel2024_LomPy} at \url{https://pypi.org/project/lompy/}. The LiDAR and cup-anemometer data analysed in this study are protected under the NearWake and DFWIND2 project framework and are not publicly available.
\paragraph{Competing interests}  We declare that the authors have no competing interests that could be perceived to influence the results or discussion reported in this paper.


\newpage
\section*{\fontsize{14}{18}\selectfont Extended data}
\setcounter{figure}{0}
\renewcommand{\figurename}{Extended Data Fig.}
\renewcommand{\tablename}{Extended Data Table}
\begin{figure}[htbp]
    \makebox[\textwidth][c]{%
        \begin{minipage}{1\linewidth}
            \centering
            \includegraphics[width=\linewidth]{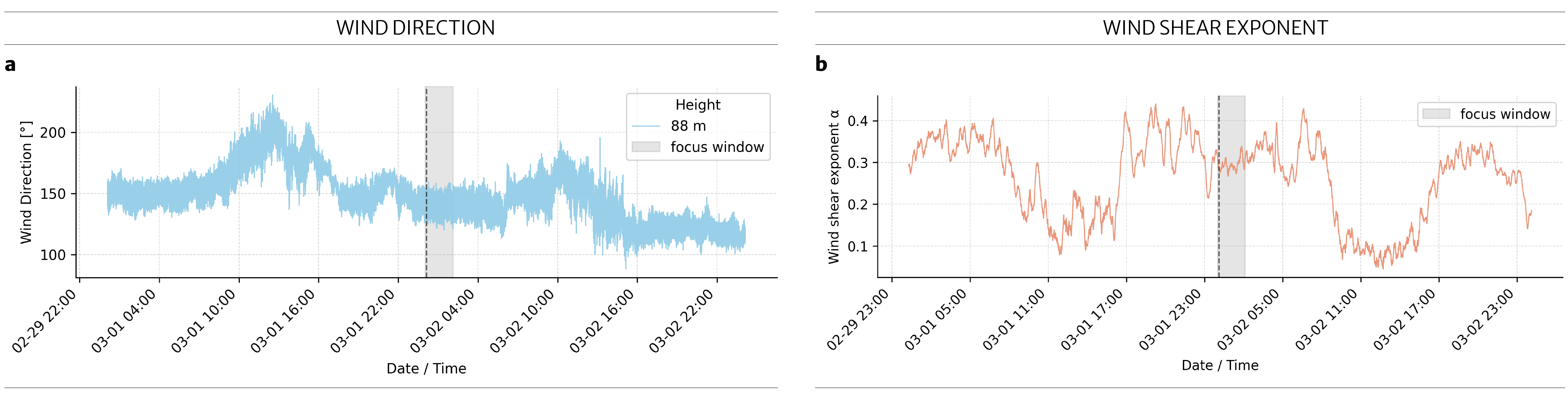}
        \end{minipage}
    }
    \caption{\captiontitle{Meteorological conditions on site}. Wind direction (a) measured by the wind vane at 88 m height. The wind shear exponent (b) was computed using wind velocities at four different heights using the cup-anemometer signals (ED Fig.\ref{fig:mfcup_48}a). The dashed line indicates the timestamp of the scan analyzed in the LiDAR-results section in Fig. \ref{fig:main_c1c2}.}
    \label{fig:ext_cup_alpha}
\end{figure}
\begin{figure}[htbp]
    \makebox[\textwidth][c]{
        \begin{minipage}{1\linewidth}
            \centering
            \includegraphics[width=\linewidth]{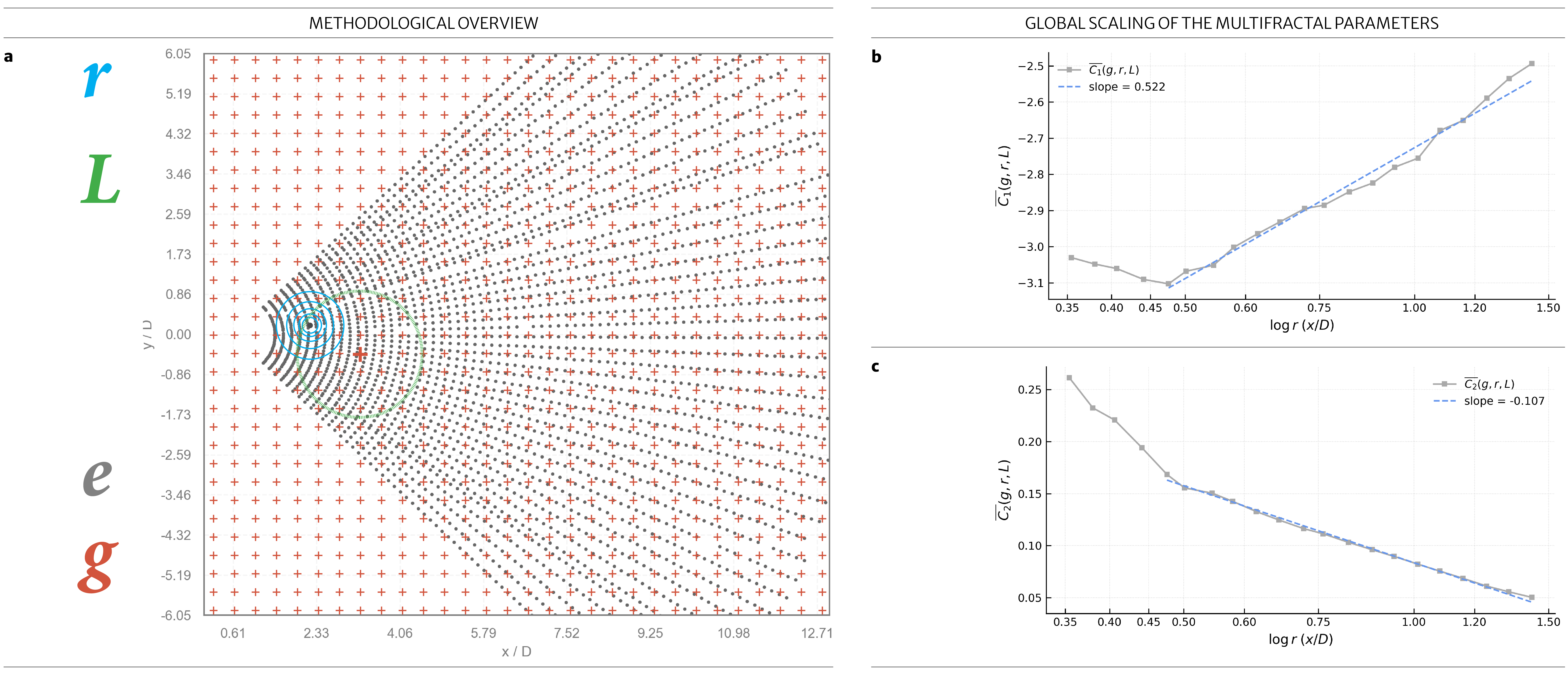}
        \end{minipage}
    }
    \caption{\captiontitle{Schematic overview of the local multifractal analysis methodology}. In panel (a), colors denote the key components of the LMF methodology: grey - original, non-homogeneously sampled LiDAR data $\kappa_e$ at points $e(x_e, y_e)$; blue - length scales $r$ used for the wavelet decomposition centred on $e$; red - estimation sites $g(x_g, y_g)$; green - extent of the local neighbourhood $L$ within which computations are performed. The globally averaged scaling, $\overline{C}_q(g,r,L) = \sum_g C_q(g,r,L)$ for order $q = 1,2$, of $C_1(g,r,L)$ (b) and $C_2(g,r,L)$ (c) is used to determine the suitable value for the local environment $L$. } 
    \label{fig:ext_method}
\end{figure} 
\begin{table}[h!]
\centering

\begin{subtable}{0.32\textwidth}
\centering
\resizebox{\linewidth}{!}{
\begin{tabular}{|c|c|c|c|c|}
\hline
 & VD(g,L) & TI(g,L) & $c_1(g,L)$ & $c_2(g,L)$ \\
\hline
VD(g,L)     & 1.00    &  &  &  \\
TI(g,L)     & $-0.7789$ & 1.00    & &  \\
$c_1(g,L)$  & $-0.4556$ & $0.3650$ & 1.00    &  \\
$c_2(g,L)$  & $-0.0232$ & $-0.0377$ & $-0.0099$ & 1.00    \\
\hline
\end{tabular}
}
\caption{Instantaneous (1 PPI scan)}
\end{subtable}
\hfill
\begin{subtable}{0.32\textwidth}
\centering
\resizebox{\linewidth}{!}{
\begin{tabular}{|c|c|c|c|c|}
\hline
 & VD(g,L) & TI(g,L) & $c_1(g,L)$ & $c_2(g,L)$ \\
\hline
VD(g,L)     & 1.00    &  &  &  \\
TI(g,L)     & $-0.8009$ & 1.00    & &  \\
$c_1(g,L)$  & $-0.4198$ & $0.4655$ & 1.00    &  \\
$c_2(g,L)$  & $-0.0331$ & $-0.1359$ & $-0.0121$ & 1.00    \\
\hline
\end{tabular}
}
\caption{Short-term (10 PPI scans)}
\end{subtable}
\hfill
\begin{subtable}{0.32\textwidth}
\centering
\resizebox{\linewidth}{!}{
\begin{tabular}{|c|c|c|c|c|}
\hline
 & VD(g,L) & TI(g,L) & $c_1(g,L)$ & $c_2(g,L)$ \\
\hline
VD(g,L)     & 1.00    &  &  &  \\
TI(g,L)     & $-0.7725$ & 1.00    & &  \\
$c_1(g,L)$  & $-0.3718$ & $0.4419$ & 1.00    &  \\
$c_2(g,L)$  & $-0.04255$ & $-0.1103$ & $-0.0309$ & 1.00    \\
\hline
\end{tabular}
}
\caption{Long-term (222 PPI scans)}
\end{subtable}

\caption{Pearson correlation between CWES and LMF metrics.}
\end{table}
\begin{figure}[htbp]
    \makebox[\textwidth][c]{%
        \begin{minipage}{1.\linewidth}
            \centering
            \includegraphics[width=\linewidth]{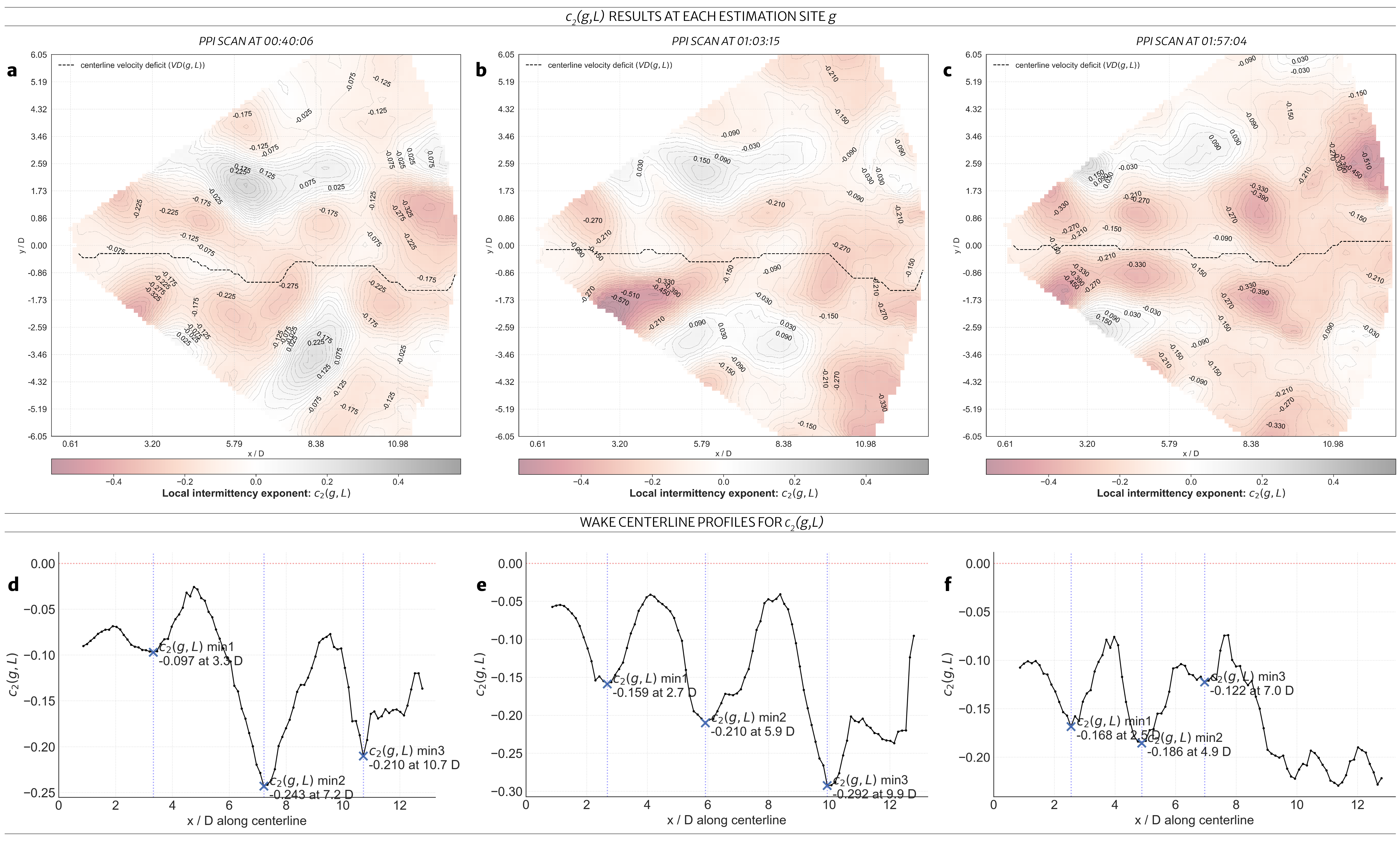}
        \end{minipage}
    }
    \caption{\captiontitle{Instantaneous results of local intermittency.} Three additional instantaneous $c_2(g,L)$ fields (a,b,c), obtained from selected PPI scans, further illustrate the recurrent nature of the intermittency cycles (d,e,f). All scans fall within the 2h (long-term) observation window. This figure complements Fig.~\ref{fig:main_c1c2}f,h in the main manuscript.}
    \label{fig:ext_int_cycles}
\end{figure}
\newpage
\begin{figure}[htbp]
    \makebox[\textwidth][c]{%
        \begin{minipage}{1.\linewidth}
            \centering
            \includegraphics[width=\linewidth]{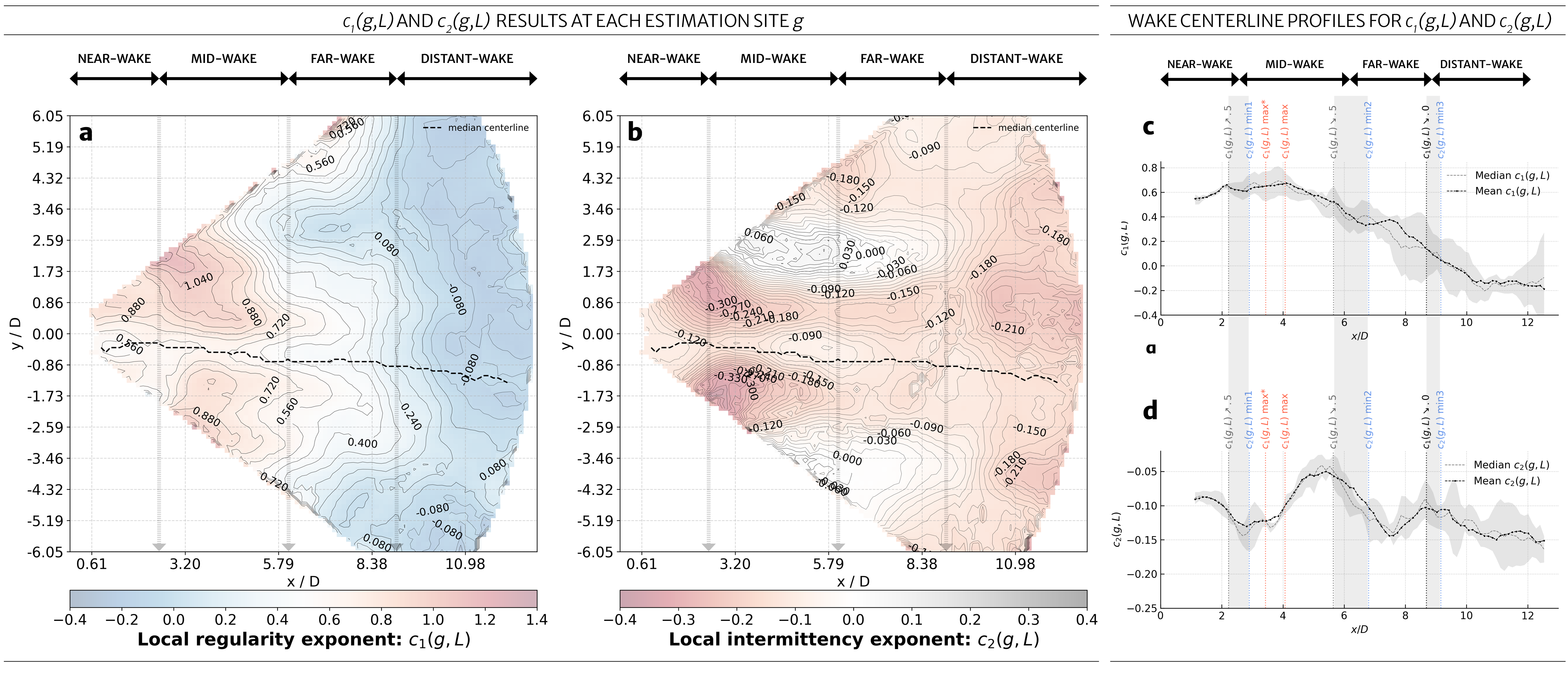}
        \end{minipage}
    }
    \caption{\captiontitle{Short-term results of the LMF metrics}. Averaged results for $c_{1}(g,L)$ (a,c) and $c_{2}(g,L)$ (b,d) across 10 PPI scans, corresponding to approximately five minutes of data. The average inflow wind speed is $6.21~\mathrm{m\,s^{-1}}$ (ED Fig.~\ref{fig:mfcup_48}.a). This figure complements Fig.~\ref{fig:short_term_results} in the main manuscript.}
    \label{fig:ext_c1c2_shortterm}
\end{figure}
\begin{figure}[h!]
    \makebox[\textwidth][c]{%
        \begin{minipage}{1.\linewidth}
            \centering
            \includegraphics[width=\linewidth]{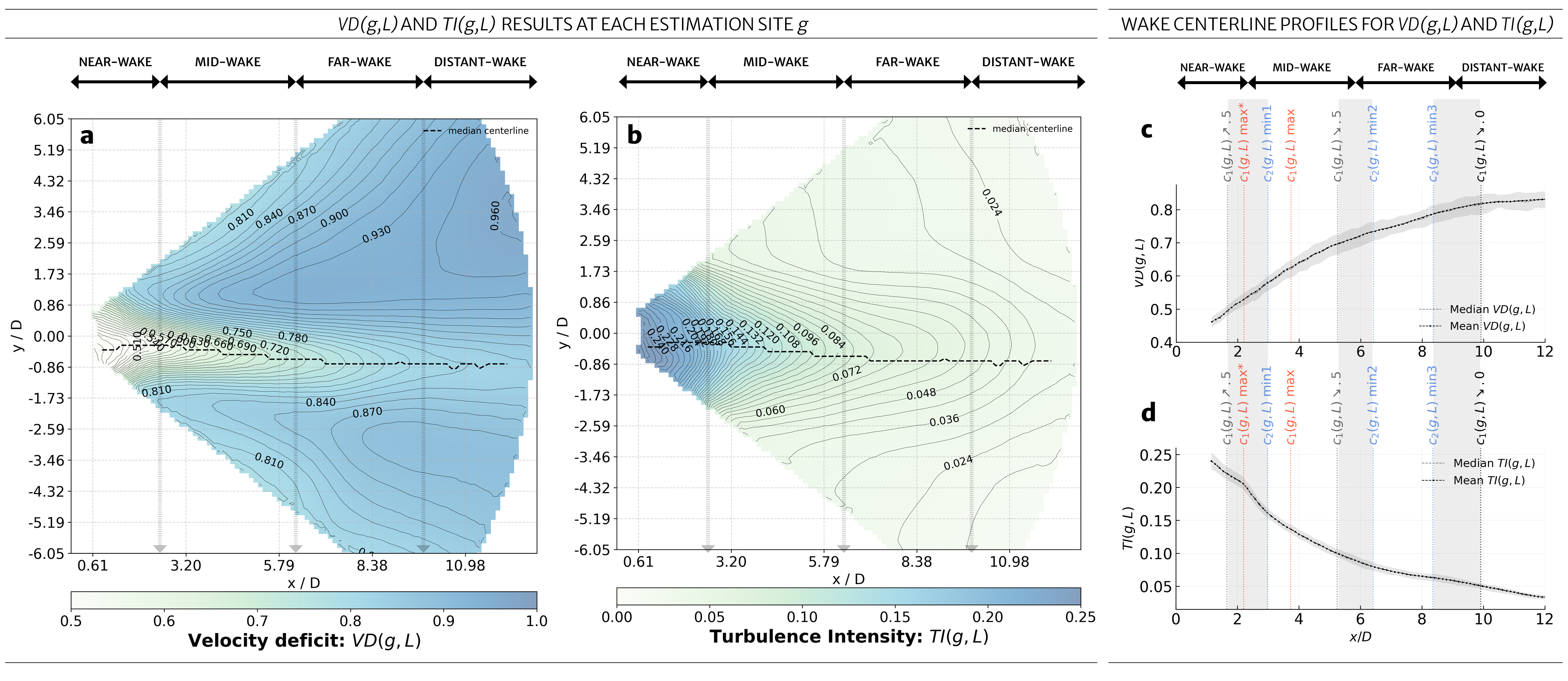}
        \end{minipage}
    }
    \caption{\captiontitle{Long-term results of the classical wind energy science metrics}. Velocity deficit $VD(g,L)$ (a) and turbulence intensity $TI(g,L)$ (b) convey distinct physical information in theory, yet in practice the two fields are highly anti-correlated (Pearson $\rho = -0.7725$). The average inflow wind speed is $6.94~\mathrm{m\,s^{-1}}$ (ED Fig.~\ref{fig:mfcup_48}.a). This figure complements Fig.~\ref{fig:main_c1c2_long} in the main manuscript.}
    \label{fig:ext_TI_VD_long}
\end{figure}
\begin{figure}[htbp]
    \makebox[\textwidth][c]{%
        \begin{minipage}{1.\linewidth}
            \centering
            \includegraphics[width=\linewidth]{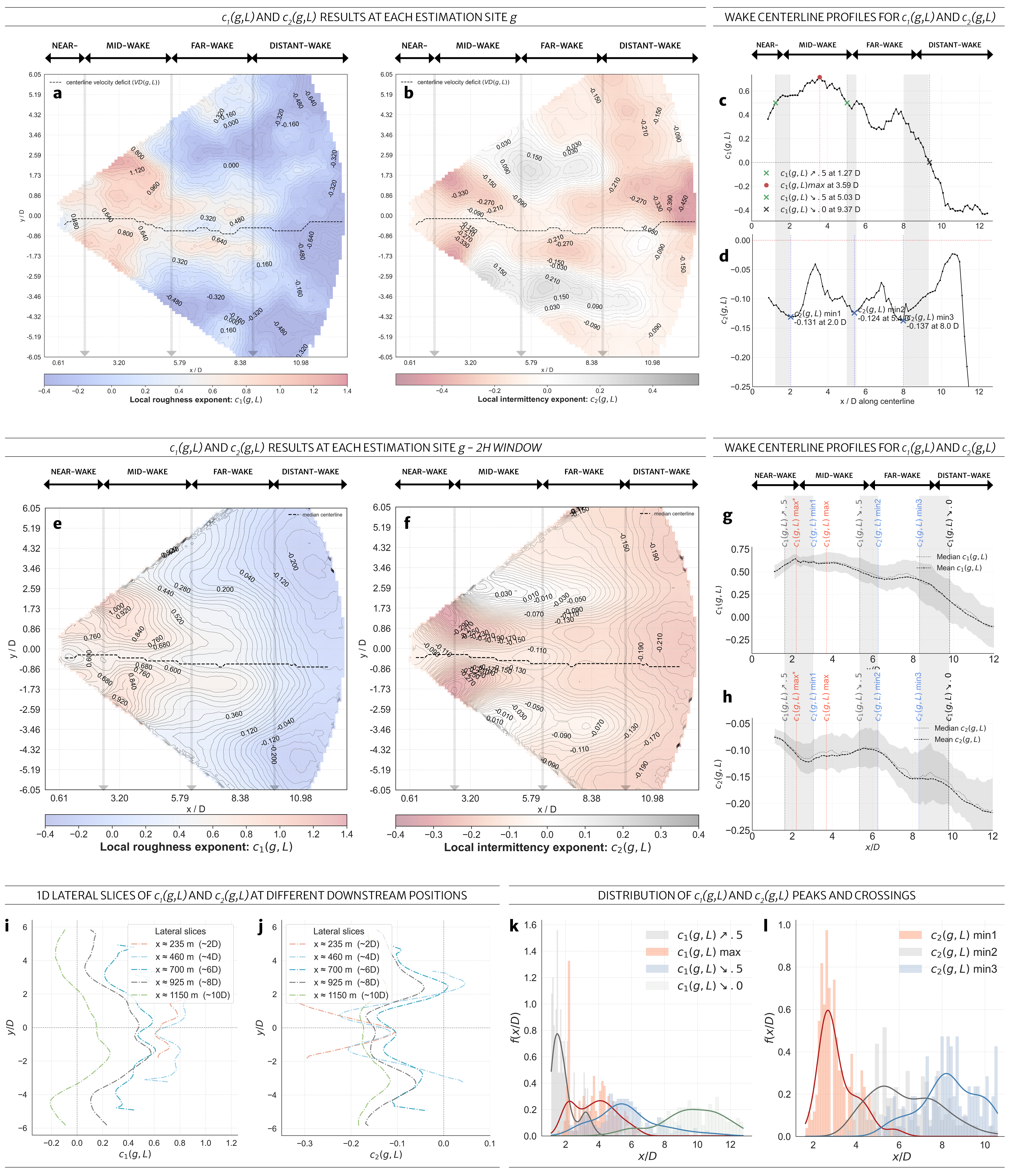}
        \end{minipage}
    }
    \caption{\captiontitle{Instantaneous and long-term results of the LMF metrics in our second case study}. Instantaneous (a,c) and averaged results for $c_{1}(g,L)$ (e,g,i,k) and instantaneous (b,d) and averaged results for $c_{2}(g,L)$ (f,h,j,l) across 222 scans, corresponding to approximately two hours of data. The inflow wind speed for panels (a-d) is $7.62~\mathrm{m\,s^{-1}}$, while the mean inflow wind speed over the 2h window for panels (e-l) is $7.45~\mathrm{m\,s^{-1}}$, as shown in ED Fig.~\ref{fig:mfcup_48}a. This second case study is included to assess the robustness of the extracted wake zones and their characteristic features reported in the main manuscript. For panels (a–d), the inflow wind speed is $1.43~\mathrm{m\,s^{-1}}$ higher than in the case study shown in Fig.~\ref{fig:main_c1c2}.
}
    \label{fig:ext_second_cs}
\end{figure}

\newpage
\begin{figure}[t!]
    \makebox[\textwidth][c]{%
        \begin{minipage}{1.0\linewidth}
            \centering
            \includegraphics[width=\linewidth]{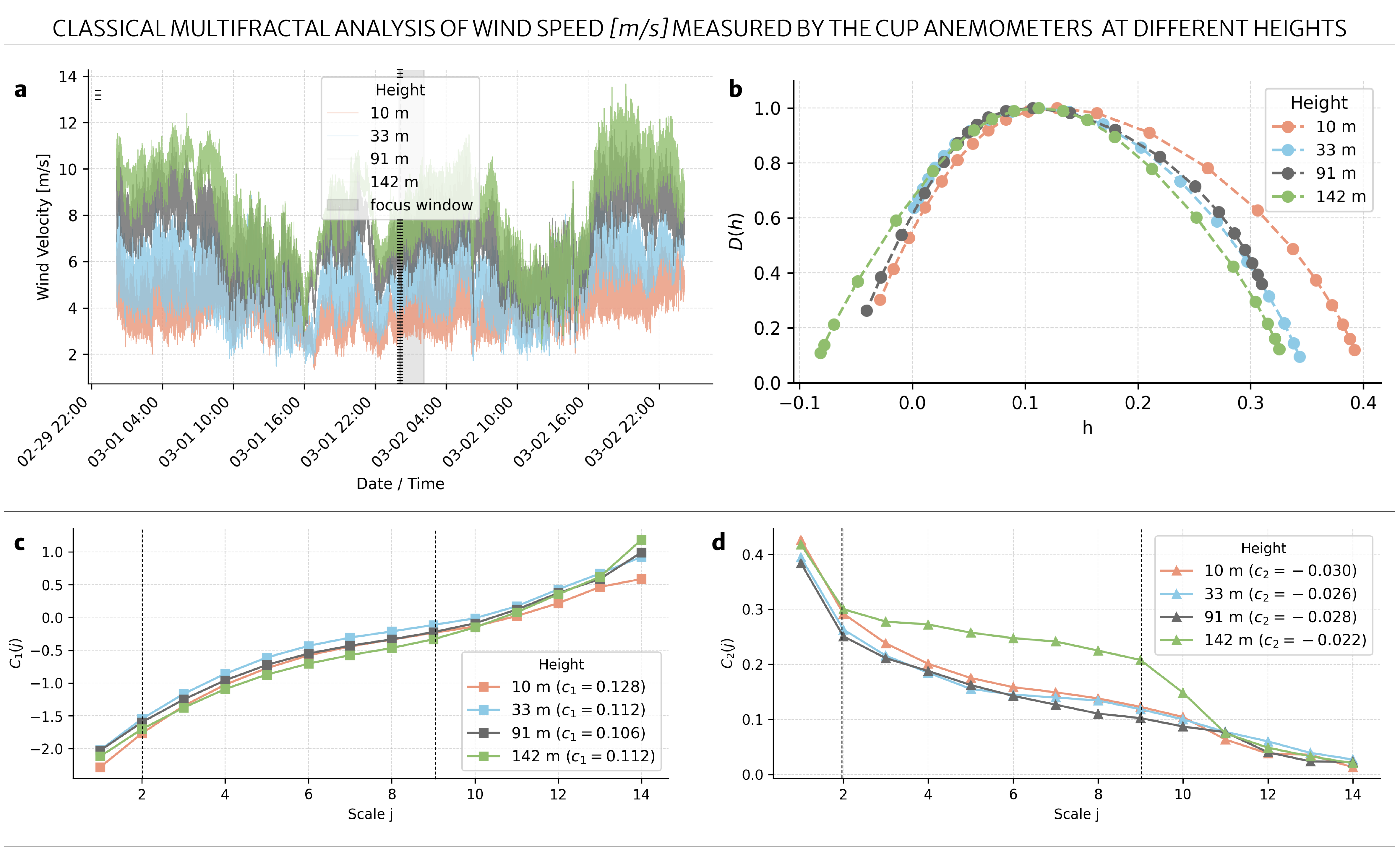}
        \end{minipage}
    }
    \caption{\captiontitle{Classical multifractal analysis - 48-hour window.} Wavelet-leader multifractal spectra (b) computed from the cup-anemometer velocity (a) signals within $\pm$24 hours of the scan time (black dashed line) in Fig.~\ref{fig:main_c1c2}. Panels (c) and (d) show the corresponding scaling of $C_1(j)$ and $C_2(j)$, respectively, along with the estimated $c_1$ and $c_2$ exponents obtained using the scale range indicated by the grey dashed lines. This figure complements Fig.~\ref{fig:mfcup_24} in the main manuscript.}
    \label{fig:mfcup_48}
\end{figure}
\begin{figure}[b!]
    \makebox[\textwidth][c]{%
        \begin{minipage}{1.\linewidth}
            \centering
            \includegraphics[width=\linewidth]{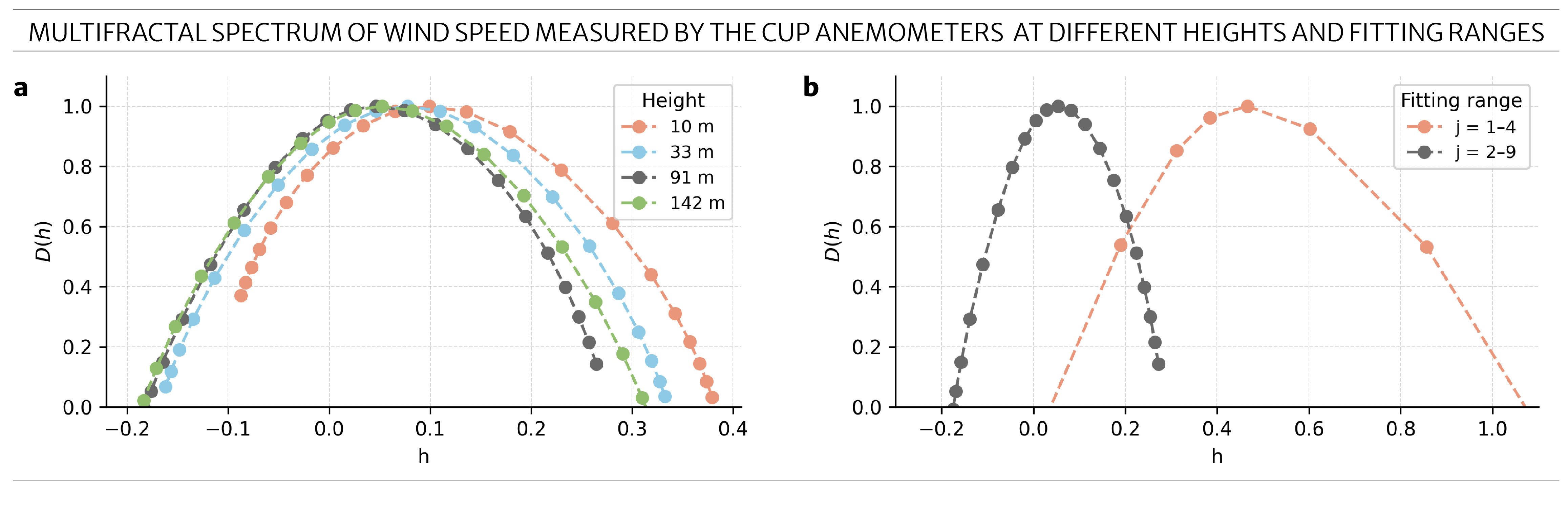}
        \end{minipage}
    }
    \caption{\captiontitle{Classical multifractal analysis - 2-hour window.} Multifractal spectra of the wind speed measured by CUP anemometers at different heights (left) and using distinct fitting ranges (right). The right panel demonstrates that restricting the regression to the smallest available scales (e.g., $j = 1\text{-}4$, red) yields a broader spectrum and an inflated roughness estimate. Extending the fit to a wider range (e.g., $j = 2\text{-}9$, grey) produces a narrower and statistically more stable spectrum, as the increased number and span of scales reduce both regression bias and the influence of small-scale noise. This results in a more robust and consistent characterization of the spectra, as a too-short fitting range is known to bias multifractal regressions, leading to inflated intermittency and unstable spectra. This figure complements Fig.~\ref{fig:mfcup_24} in the main manuscript.}
    \label{fig:ext_appendix_mf_compare}
\end{figure}

\begin{figure}[htbp]
    \makebox[\textwidth][c]{%
        \begin{minipage}{.85\linewidth}
            \centering
            \includegraphics[width=\linewidth]{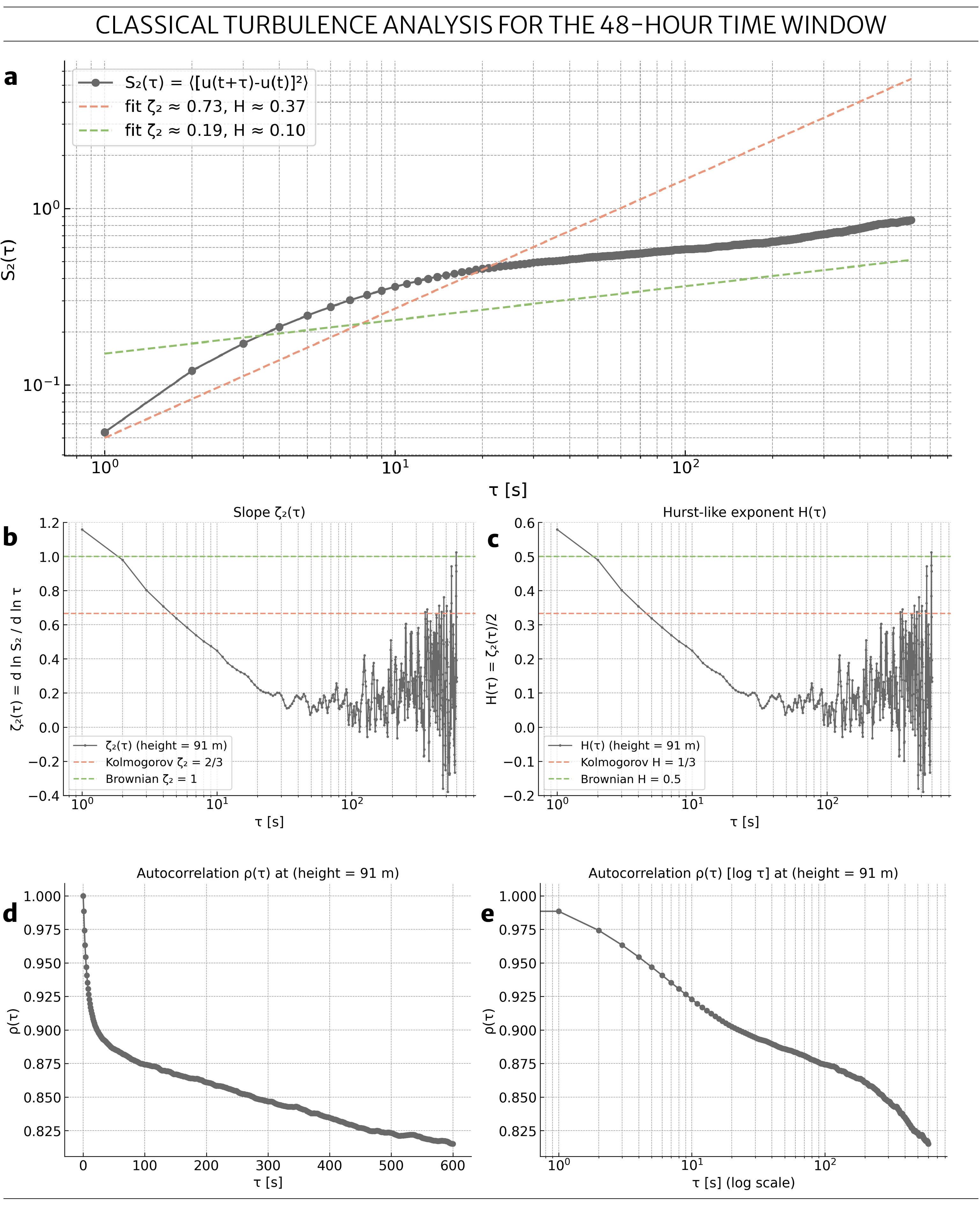}
        \end{minipage}
    }
    \caption{\captiontitle{Increments-based analysis of the cup-anemometer velocity signal at hub height.} The second-order structure function (a) exhibits two characteristic scaling ranges: one up to roughly $1-8$ s, corresponding broadly to Kolmogorov scaling, and a second - more dominant - scaling between $8–600$ s that shows rougher or shorter-correlated spatial dependencies with a Hurst exponent of $0.10$. Panel (b,c) shows the local slope $\zeta_{2}(\tau)$, obtained as the logarithmic derivative of the second-order structure function. The absence of a clear plateau across scales indicates that no well-defined inertial range is present in this 48-hour period. Panel (d,e) displays the autocorrelation function $\rho(\tau)$, which decays only slowly with increasing time lag and does not approach zero over the available temporal length scales. This figure complements Fig.~\ref{fig:mfcup_24} in the main manuscript.}
    \label{fig:ext_turb}
\end{figure}

\end{document}